\documentclass[twocolumn,prl,superscriptaddress]{revtex4-1}
\usepackage{graphicx}
\usepackage{amssymb,amsmath}
\usepackage{bm}
\usepackage{dcolumn}
\usepackage{subfigure}
\usepackage{float}
\usepackage{url}
\usepackage{xcolor}
\usepackage{ulem}
\usepackage{sidecap}
\usepackage{bbold}
\allowdisplaybreaks

\begin{document}

\title{Odd-parity topological superfluidity for fermions in a bond-centered square optical lattice}
\author{Zhi-Fang Xu}
\email{zfxu83@gmail.com}
\affiliation{Institute for Quantum Science and Engineering and Department of Physics, Southern University of Science and Technology of China, Shenzhen 518055, China}
\affiliation{MOE Key Laboratory of Fundamental Physical Quantities Measurements, School of Physics, Huazhong University of Science and Technology, Wuhan 430074, China}
\author{Andreas Hemmerich}
\affiliation{Institut f\"ur Laser-Physik, Universit\"at Hamburg, Luruper Chaussee 149, 22761 Hamburg, Germany}
\author{W. Vincent Liu}
\email{wvliu@pitt.edu}
\affiliation{Department of Physics and Astronomy, University of Pittsburgh, Pittsburgh, Pennsylvania 15260, USA}
{\color{red} }

\begin{abstract}
  We propose a physical scheme for the realization of two-dimensional topological odd-parity superfluidity in a spin-independent bond-centered square optical lattice based upon interband fermion pairing. The $D_4$ point-group symmetry of the lattice protects a quadratic band crossing, which allows one to prepare a Fermi surface of spin-up fermions with odd parity close to the degeneracy point. In the presence of spin-down fermions with even parity populating a different energetically well separated band, odd-parity pairing is favored. Strikingly, as a necessary prerequisite for pairing both Fermi surfaces can be tuned to match well. As a result, topological superfluid phases emerge in the presence of merely s-wave interaction. Due to the $Z_2$ symmetry of these odd-parity superfluids, we infer their topological features simply from the symmetry and the Fermi-surface topology as confirmed numerically.
\end{abstract}

\maketitle

{\it Introduction.}---
Topological superconductivity and its charge neutral analogue of topological superfluidity, are intriguing forms of topological matter, long sought after in electronic or cold atomic systems~\cite{Volovik2003, Read2000, Nayak2008, Kallin2012, Sato2016, Gurarie2007, Zhang2008, Sato2009, Cooper2009, Liu2014XJL, Buhler2014, Liu2014, Zhang2015, Wang2016, Wu2016}. Two approaches to topological superconductivity have been taken, either using intrinsic topological superconductors or hetero-structures, for example, made of an $s$-wave superconductor and a topological insulator~\cite{Fu2008, Jia2012, Jia2015}. As an example of a homogeneous system, strontium ruthenate~\cite{Maeno1994, Mackenzie2003} has been widely discussed as a possible candidate for a topological chiral $p_x+ip_y$ superconducting phase. The evidence, however, has remained inconclusive. A powerful alternative route towards homogeneous systems showing topological superconductivity is the use of cold atoms~\cite{Zhang2008, Sato2009, Cooper2009, Liu2014XJL, Buhler2014, Liu2014,Zhang2015, Wang2016, Wu2016}. In many studies, an interaction involving higher partial waves is required, e.g., a $p$-wave interaction induced by spin-orbital coupling. In other cold atom studies this experimentally demanding constraint was relaxed. For instance, it was demonstrated as a proof of principle that pairing fermionic atoms from $s$ and $p$ orbital bands by $s$-wave interaction may give rise to the possibility of a topological chiral $p$-wave superfluid if the two spin components are loaded into different optical sublattices~\cite{Liu2014}. The realization of the required spin dependence of the optical lattice potential, however, poses another significant experimental challenge. Particularly, for the widely used fermionic Lithium atoms, the small fine structure splitting practically rules out the possibility of spin-dependent lattices without running into substantial heating.

In this Letter, we show that topological superfluidity can naturally emerge in a spin-1/2 Fermi gas inside an optical lattice by pairing orbital states of odd and even parities. Our model bypasses the notorious technical complexities that have impeded experiments to date, such as the necessity to engineer synthetic gauge fields spin-dependent optical lattice potentials, and higher partial-wave interaction. The two-dimensional (2D) spin-independent optical lattice is derived from a single monochromatic laser beam, and provides a $D_4$ point-group symmetry and a band structure with a quadratic band degeneracy point protected by odd parity. Cooper pair formation only requires $s$-wave on-site interactions. Here we summarize the main features and results of our model. 1.~The energy spectrum of the non-interacting part of the model is characterized by two adjacent (2nd and 3rd) energy bands that are both convex. This is in contrast to an earlier study of an inter-band pairing mechanism~\cite{Liu2014} where two adjacent bands generally possess curvatures of opposite sign at the relevant high symmetry points. We shall elaborate later that this is very important to well match the Fermi surfaces of the two spin species residing in these two bands. 2.~The fermionic states close to the Fermi surfaces are mainly composed of highly overlapping orbitals with opposite parities, where the appearance of odd parity orbitals is guaranteed by tuning the Fermi surface of one spin-component close to the quadratic band crossing. 3.~In this condition, our calculation shows that the components of the superfluid order parameter are made from pairings of $s$-$p$ and $p$-$d$ orbitals. They have odd parity and spontaneously break time-reversal symmetry, thus realizing an odd-parity topological chiral  superfluid. Calculation of gapless chiral edge modes further support and classify the  topological nature of the phase.

\begin{figure}
\centering
\includegraphics[width=\linewidth]{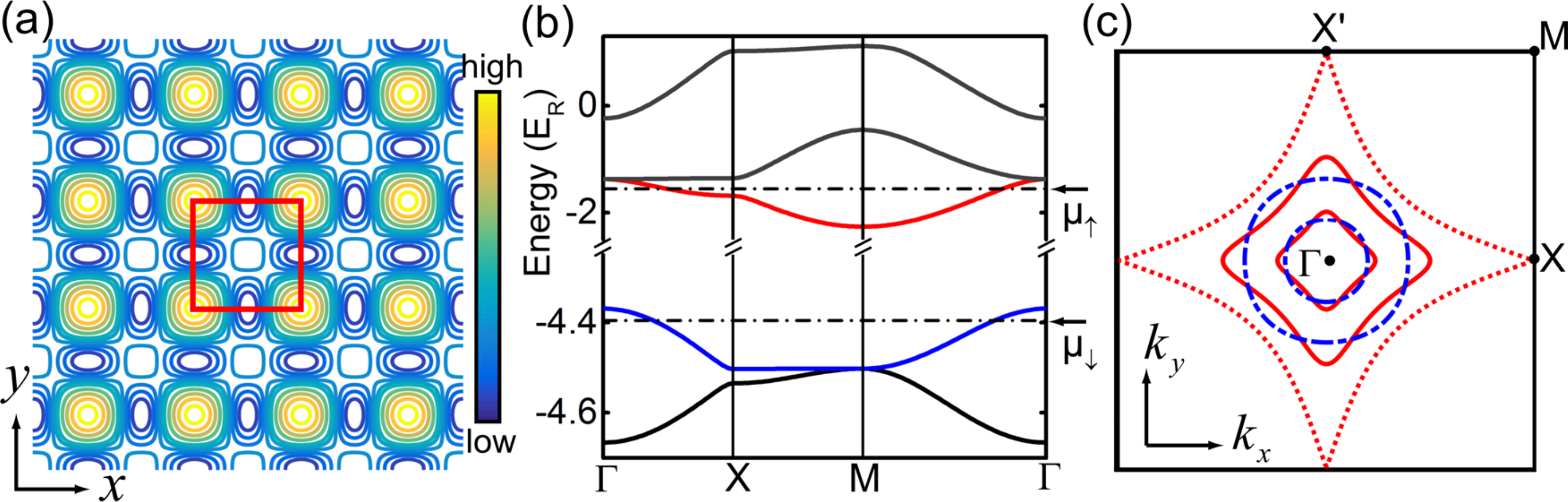}
\caption{ (color online). (a) Contour plot of the optical lattice potential $V(x,y)$ with $V_1=V_2=5 E_R$, where $E_R=h^2/(4ma^2)$ is the recoil energy and $h$ is the Planck constant. The red solid square denotes a unit cell of the lattice. (b) Single-particle energy bands along high-symmetry lines. Dotted-dash lines denote the spin-up and spin-down chemical potentials. (c) Fermi surfaces for the spin-up (red solid line) and spin-down (blue dotted-dash line) components. From inner to outer, $31/32$ and $7/8$ of the 2nd (3rd) band are filled by the spin-up (spin-down) fermions for the normal state. Red dotted line separates two cases where the spin-up Fermi surface encloses the $\Gamma$ point and the M point, respectively.}
\label{fig1}
\end{figure}

{\it Model.}---
In cold atom experiments a great variety of unconventional optical lattice geometries can be readily implemented such as  honeycomb~\cite{Sengstock2010}, Kagome~\cite{Jo2012}, Lieb~\cite{Taiee2015}, and checkerboard~\cite{Sebby-Strabley2006, Wirth2011} lattices. Here, we focus on the 2D optical lattice geometry discussed in Ref.~\cite{Sun2012} with the potential
\begin{eqnarray}
  V(x,y)&=&-V_1[\cos(k_Lx)+\cos(k_Ly)]\nonumber \\
  &+&V_2[\cos(k_Lx+k_Ly)+\cos(k_Lx-k_Ly)],
  \label{potential}
\end{eqnarray}
illustrated in Fig.~\ref{fig1}(a). Here, $a=2\pi/k_L$ is the lattice constant. Due to the hybridization among orbitals with different parities and the associated $D_4$ point-group symmetry, a quadratic band crossing point (QBCP) appears at the $\Gamma$ point between the 3rd and the 4th energy bands. With arbitrary weak short-range repulsive interaction, the QBCP is unstable towards the formation of topological states, e.g. a quantum anomalous Hall phase~\cite{Sun2009}. It has been pointed out in Ref.~\cite{Sun2012} that for $V_2/V_1>1/2$ the potential of Eq.~(\ref{potential}) can be readily formed using a single monochromatic laser beam.

In this Letter, we study a Fermi gas prepared in the optical lattice potential of Eq.~(\ref{potential}). In contrast to Ref.~\cite{Sun2012} we here consider three new aspects: a spin population imbalance~\cite{Zwierlein2006, Partridge2006} such that the Fermi-surfaces of spin-up and spin-down components intersect different bands, attractive rather than repulsive $s$-wave interaction, and a modified band structure resulting from a different choice of the parameter ratio $V_2/V_1$. For the case of $V_2/V_1\sim1$, the 2nd band is shifted downwards and separated from the 3rd and 4th bands, while the QBCP at the $\Gamma$ point is retained, as illustrated in Fig.~\ref{fig1}(b). Remarkably, both the 2nd and 3rd bands are convex, with their band minima (maxima) both being located at or near the M ($\Gamma$) point. This feature is important for Cooper pairing. By tuning the spin-up (down) Fermi surface crossing the 3rd (2nd) band,  we are able to obtain well-matched Fermi surfaces, as demonstrated in Fig.~\ref{fig1}(c). In contrast to the case of equal spin population, for which both theoretical and experimental studies agree on conventional $s$-wave pairing~\cite{Chin2006, Zhai2007, Moon2007}, we find that appropriate tuning of the spin imbalance can result in the emergence of chiral odd-parity pairing, to be discussed next.

Including the attractive contact interaction, the quasi 2D system is well described by the Hamiltonian
\begin{eqnarray}
  \hat{H}=\int d\mathbf{r}\bigg[\sum\limits_{\sigma=\uparrow,\downarrow}
  \hat{\psi}_{\mathbf{r}\sigma}^{\dag}(H_0-\mu_{\sigma})\hat{\psi}_{\mathbf{r}\sigma}
  -U\hat{\psi}_{\mathbf{r}\uparrow}^{\dagger}\hat{\psi}_{\mathbf{r}\downarrow}^{\dag} \hat{\psi}_{\mathbf{r}\downarrow}\hat{\psi}_{\mathbf{r}\uparrow}\bigg],
  \label{hamiltonian}
\end{eqnarray}
where $\mathbf{r}=(x,y)$, $H_{0}=-\hbar^2(\partial_x^2+\partial_y^2)/2M+V(x,y)$, $\mu_{\sigma}$ being the chemical potential, and $U>0$. In the mean-field framework, we consider only the on-site fermion pairing and define the order parameter as
\begin{eqnarray}
  \Delta_{\mathbf{r}}\equiv-U\langle \hat{\psi}_{\mathbf{r}\downarrow}\hat{\psi}_{\mathbf{r}\uparrow}\rangle.
  \label{orderparamter}
\end{eqnarray}
Then, the interaction part Hamiltonian becomes
$\hat{H}_{\rm int}=\int d\mathbf{r}\Big(\hat{\psi}_{\mathbf{r}\uparrow}^{\dag} \hat{\psi}^{\dag}_{\mathbf{r}\downarrow}\Delta_{\mathbf{r}}+\rm h.c.\Big)+\int d\mathbf{r}|\Delta_{\mathbf{r}}|^2/U$. In the following, we shall focus on the case of  well-matched spin-up and down Fermi surfaces by tuning the chemical potentials as shown in Fig.~\ref{fig1}(c). Therefore, the possibility of Fulde-Ferrell-Larkin-Ovchinnikov (FFLO) states~\cite{Fulde1964,Larkin1964,Liao2010} is suppressed. It is then reasonable to consider the conventional BCS pairing and assume that the order parameter takes the same periodicity as the optical lattice potential $V$. This leads to $\Delta_{\mathbf{r}}=\sum_{\mathbf{Q}}\Delta_{\mathbf{Q}}\exp[i\mathbf{Q}\cdot\mathbf{r}]$, where $\mathbf{Q}$ is the reciprocal lattice vector. To diagonalize the Hamiltonian, we expand the field operator by the Bloch waves as $\hat{\psi}_{\mathbf{r}\sigma}=\sum_{n\mathbf{k}}\phi_{n\mathbf{k}}(\mathbf{r})\hat{\psi}_{n\mathbf{k}\sigma}$, where $n$ denotes the band index and the Bloch waves can be further expanded by the plane-wave basis as $\phi_{n\mathbf{k}}=\frac{1}{\sqrt{\mathcal{V}}}\sum_{\mathbf{Q}} \varphi_{n\mathbf{k}}(\mathbf{Q})\exp[i(\mathbf{k}+\mathbf{Q})\cdot\mathbf{r}]$ with $\mathcal{V}$ being the system volume. Thus, the mean-field Hamiltonian is given by
\begin{eqnarray}
 \hat{H}_{\rm MF}
 &=&\sum\limits_{n\mathbf{k}\sigma}[\xi_{n}(\mathbf{k})-\mu_{\sigma}] \hat{\psi}_{n\mathbf{k}\sigma}^{\dag}\hat{\psi}_{n\mathbf{k}\sigma}
 +\frac{\mathcal{V}}{U}\sum\limits_{\mathbf{Q}}|\Delta_{\mathbf{Q}}|^2
 \nonumber\\
 &+&\sum\limits_{nm\mathbf{k}}\Big(\hat{\psi}_{n\mathbf{k}\uparrow}^{\dag} \hat{\psi}_{m,-\mathbf{k}\downarrow}^{\dag}\Delta_{nm\mathbf{k}}+\rm h.c.\Big),
\end{eqnarray}
where $\xi_{n}(\mathbf{k})$ is the single-particle energy of the $n$-th band at the crystal momentum $\mathbf{k}$, $\Delta_{mn\mathbf{k}}= \sum_{\mathbf{Q}}\Delta_{\mathbf{Q}}(M_{mn\mathbf{k}}^{\mathbf{Q}})^*$,  $M_{mn\mathbf{k}}^{\mathbf{Q}}=\sum_{\mathbf{K}} \varphi_{m,-\mathbf{k}}(-\mathbf{K})\varphi_{n\mathbf{k}}(\mathbf{K}+\mathbf{Q})$, and $\mathbf{K}$ is the reciprocal lattice vector.
Numerically, we take into account the lowest four bands, as they deviate from the upper bands. To obtain the ground state, we use the simulated annealing method to find the global minimum of the grand potential $\Omega\equiv -\frac{1}{\beta}\ln\text{Tr}\exp[-\beta\hat{H}_{\rm MF}]$, accompanied by the self-consistent iteration method~\cite{suppl}.

\begin{figure}[tbp]
\centering
\includegraphics[width=0.9\linewidth]{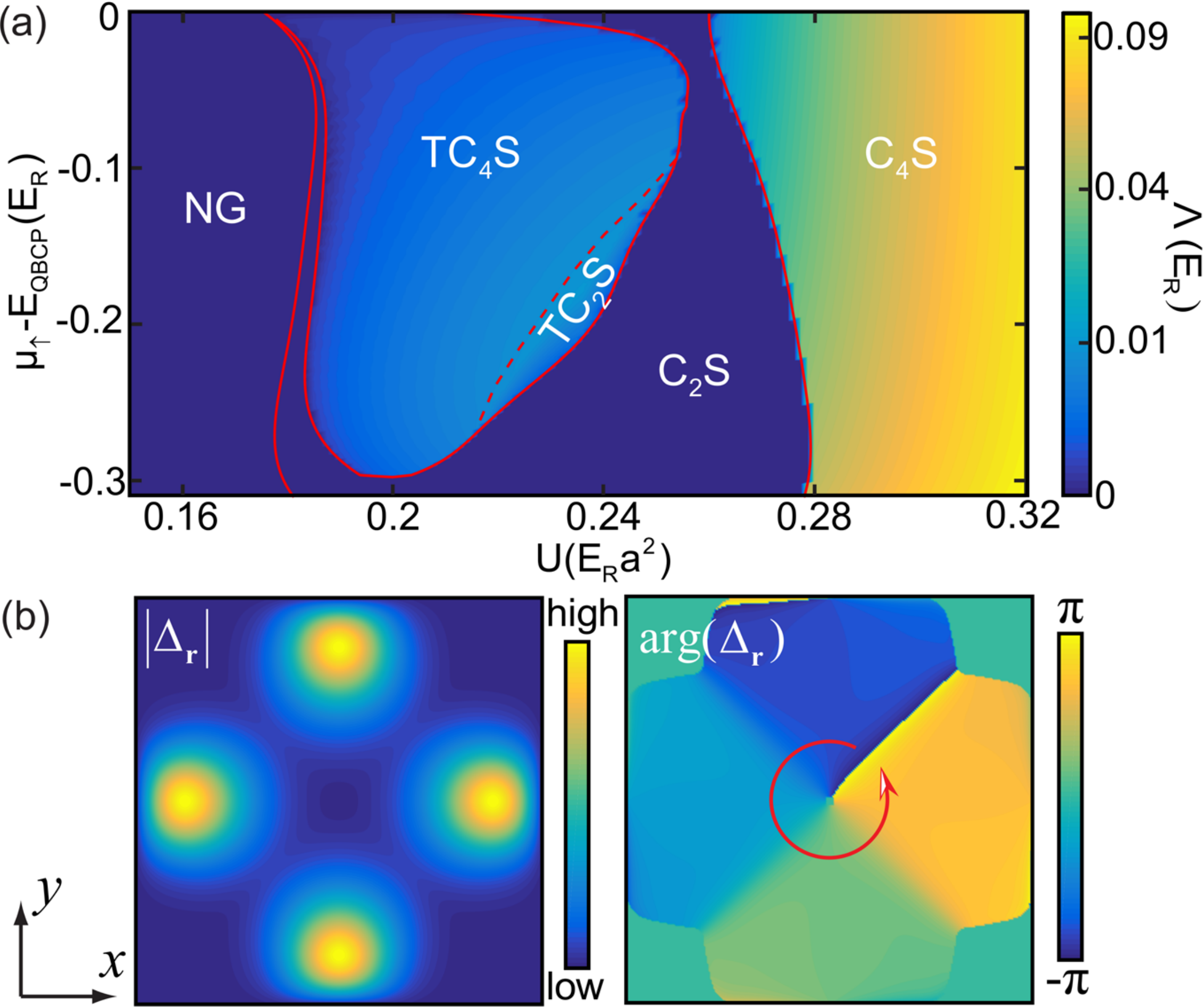}
\caption{ (color online). (a) Zero-temperature ground-state phase diagram by varying the $s$-wave interaction and the chemical potentials. Spin-up and spin-down chemical potentials are changed simultaneously to make the enclosed area for two Fermi surfaces equal in the momentum space and $\mu_{\uparrow}\in [E_{\rm SEP}, E_{\rm QBCP}]$, where $E_{\rm QBCP}$ denotes the single-particle energy at the quadratic band crossing and when $\mu_{\uparrow}=E_{\rm SEP}$, the spin-up Fermi surface is denoted by the red dotted line in Fig.~\ref{fig1}(c). Five different phases are denoted by NG, $\rm TC_4S$, $\rm TC_2S$, $\rm C_4S$, and $\rm C_2S$, respectively. The quasiparticle excitation gap $\Lambda$ is shown according to the color gauge. (b) The characteristic spatial distribution of the order parameter in one unit cell for the superfluid phase preserving the four-fold rotation symmetry.}
\label{fig2}
\end{figure}

{\it Topological odd-parity superfluid.}---Before showing the ground states, we first analyze the underlying symmetry of the interband pairing for the spin-imbalanced system. To provide a more intuitive picture, we consider two different tight-binding (TB) models~\cite{suppl} to describe the lowest four bands obtained from a numerical plane-wave expansion as shown in Fig.~\ref{fig1}(b). In Ref.~\cite{suppl} we point out that both TB models describe the numerically determined band structure with excellent precision (see Fig.~S3 of) and generate the same phases when the attractive interaction is turned on. Here, as one example, we consider a TB model involving four orbitals $s$, $p_x$, $p_y$, and $d_{x^2-y^2}$ centered at the center of the unit cell denoted by the red square shown in Fig.~\ref{fig1}(a). Their corresponding annihilation operators are $\hat{s}$, $\hat{p}_x$, $\hat{p_y}$, and $\hat{d}$.

When the spin-up Fermi surface is tuned close to the degenerate point ($\Gamma$ point) between the 3rd and 4th bands. The spin-up fermions close to the Fermi surface are mainly composed by the odd-parity $p$ orbitals~\cite{Sun2012,suppl}. In contrast, close to the spin-down Fermi surface which is tuned to lie near the maximum of the 2nd band, the fermions are mainly composed of even-parity orbitals. All these features can be readily confirmed by diagonalizing the single-particle Hamiltonian~\cite{suppl}. In the weak-coupling limit, the pairing is mainly among fermions close to the Fermi surfaces, and hence Cooper pairing takes place mainly between odd-parity spin-up fermions and even-parity spin-down fermions, leading to odd-parity superfluidity. From a symmetry point of view, the pairing order parameter may largely inherit the $D_4$ point-group symmetry of the system. The maximally symmetric pairing phase corresponds to locking the phase difference between two degenerate odd-parity orbitals at $\pm \pi/2$ during pairing, which leads to $\langle \hat{d}_{\downarrow}\hat{p}_{x,\uparrow}\rangle=\mp i\langle \hat{d}_{\downarrow}\hat{p}_{y,\uparrow}\rangle$ and $\langle \hat{s}_{\downarrow}\hat{p}_{x,\uparrow}\rangle=\pm i\langle \hat{s}_{\downarrow}\hat{p}_{y,\uparrow}\rangle$.

Our numerics confirms the existence of the anticipated maximally symmetric odd-parity superfluid phases which are invariant under the combined $\pm \pi/2$ gauge rotation and the $C_4$ space rotation symmetry. The corresponding order parameter are shown in Fig.~\ref{fig2}(b). In addition, we find other phases with lower symmetries.  Figure~\ref{fig2}(a) shows the zero-temperature ground-state phase diagram calculated by the plane-wave expansion. To facilitate fermion pairing, spin-up and -down chemical potentials are adjusted simultaneously to match the enclosed area of the two Fermi surfaces in momentum space. By varying the value of $s$-wave contact interaction between two spin components  and the chemical potentials, we find five different phases.

Because the two Fermi surfaces cross different bands, a finite interaction is needed to get into the superfluid phase. When the interaction is too weak, only a normal gas (NG) phase is obtained. Increasing the interaction gives rise to four different superfluids with odd parity. Two superfluid phases denoted by $\rm TC_4S$ and $\rm TC_2S$ are topologically nontrivial while the others are topologically trivial. Two phases denoted by $\rm TC_4S$ and $\rm C_4S$, spontaneously break the time-reversal symmetry but preserve the $C_4$ rotation symmetry, and are accompanied by a full bulk gap close to the zero energy. We find that both show nonzero orbital angular momenta for the center-of-mass motion of paired fermions, as illustrated in Fig.~\ref{fig2}(b) where vortices are found present in each unit cell for both states. This feature is reminiscent of the interaction-driven bosonic chiral superfluid in a checkerboard lattice studied in Ref.~\cite{Xu2016}. The other two phases, denoted by $\rm TC_2S$ and $\rm C_2S$, preserve only the $C_2$ rotation symmetry. The difference between them is that the $\rm TC_2S$ phase also breaks the time-reversal symmetry and shows a full bulk gap, while the $\rm C_2S$ phase preserves the time-reversal symmetry similar to the conventional $p$-wave superfluid with a real pairing order parameter and supports gapless excitations.

To determine the topological behavior of the odd-parity superfluid phases, we can rely on the criterion discussed in Refs.~\cite{Fu2010,Sato2010}, where the authors show that the topology of the full-gapped odd-parity superconductor -- with or without the time-reversal symmetry -- can be simply inferred from the Fermi-surface topology, e.g. the number of the time-reversal invariant (TRI) momenta enclosed by the Fermi surface. For the spin-imbalanced system we discussed, the Bogoliubov-de Gennes (BdG) Hamiltonian is given by~\cite{suppl}
\begin{eqnarray}
    \mathcal{H}_{\rm BdG}(\mathbf{k})=
    \left(
    \begin{array}{cc}
    H_0(\mathbf{k})-\mu_{\uparrow}\mathbb{1} & \hat{\Delta}(\mathbf{k}) \\
    \hat{\Delta}^{\dagger}(\mathbf{k}) & -H_0(\mathbf{k})+\mu_{\downarrow}\mathbb{1} \\
    \end{array}
    \right),
\label{bdgHamiltonian}
\end{eqnarray}
where $H_0(\mathbf{k})$ is a diagonal matrix with elements $[H_0(\mathbf{k})]_{nn}=\xi_{n}(\mathbf{k})$ and $[\hat{\Delta}(\mathbf{k})]_{nm}=\Delta_{nm\mathbf{k}}$.

As the lattice potential of Eq.~(\ref{potential}) is invariant under the $D_4$ symmetry group, the single-particle band structure exhibits the inversion symmetry $PH_0(\mathbf{k})P=H_0(-\mathbf{k})$, where $P$ is an inversion operator with inversion center defined at the center of the unit cell denoted by the red square shown in Fig.~\ref{fig1}(a). Focusing on the odd-parity superfluids shown in Fig.~\ref{fig2}(a), the order parameter satisfies $\Delta_{\mathbf{r}}=-\Delta_{-\mathbf{r}}$. Further choosing specific relative global phases for the Bloch waves at opposite momenta when calculating $M_{mn\mathbf{k}}^{\mathbf{Q}}$, we could make $\Delta_{mn\mathbf{k}}=-\Delta_{mn,-\mathbf{k}}$, leading to $P\hat{\Delta}(\mathbf{k})P=-\hat{\Delta}(-\mathbf{k})$. We thus find that the BdG Hamiltonian for the odd-parity superfluid has a $Z_2$ symmetry
\begin{eqnarray}
\tilde{P}\mathcal{H}_{\rm BdG}(\mathbf{k})\tilde{P}=\mathcal{H}_{\rm BdG}(-\mathbf{k}),\quad \tilde{P}\equiv P\tau_z,
\end{eqnarray}
where $\tau_z$ is a diagnoal matrix with diagonal elements $[\mathbb{1},-\mathbb{1}]$.
With this $Z_2$ symmetry, we can define a $Z_2$ invariant $\nu$ for characterizing the topology of the superfluid~\cite{Fu2010,Sato2010}:
\begin{eqnarray}
(-1)^{\nu}=\prod\limits_{\alpha,\ell}\pi_{\ell}(\Gamma_{\alpha}).
\end{eqnarray}
where $\pi_{\ell}(\Gamma_{\alpha})$ and $\mathcal{E}_{\ell}(\Gamma_{\alpha})$ are the eigenvalues of the $\tilde{P}$ and $\mathcal{H}_{\rm BdG}(\Gamma_{\alpha})$ on their common eigenstates at TRI points $\Gamma_{\alpha}$ and the product over $\ell$ includes quasiparticle excitations with $\mathcal{E}_{\ell}(\Gamma_{\alpha})<0$. In the weak-coupling limit, the quasiparticle eigenstates can be approximated by Bloch states of $H_0$~\cite{Fu2010,Sato2010}. We thus find
\begin{eqnarray}
(-1)^{\nu}=\prod\limits_{\alpha,n}\bar{p}_{n}(\Gamma_{\alpha})
\text{sgn}[\mu_{\downarrow}-\xi_{n}(\Gamma_{\alpha})],
\end{eqnarray}
where the product over $n$ covers all bands of $H_0$ and  $\bar{p}_n(\Gamma_{\alpha})=p_{n}(\Gamma_{\alpha})$ if $[\mu_{\uparrow}-\xi_n(\Gamma_{\alpha})][\xi_n(\Gamma_{\alpha})-\mu_{\downarrow})]>0$ and $1$ otherwise. Here $p_{n}(\Gamma_{\alpha})$ is the eigenvalue of the parity operator for the $n$-th Bloch states at $\Gamma_{\alpha}$. There is a crucial difference from the $Z_2$ invariant defined in the context of electronic superconductivity~\cite{Fu2010,Sato2010}. Here, due to spin population imbalance,  the summation of the occupied spin-up bands and the unoccupied spin-down bands overcompletely covers the complete set of single-particle energy bands. This requires us to consider the parity for the states in between the two Fermi surfaces at TRI points.

\begin{figure}[tbp]
\centering
\includegraphics[width=\linewidth]{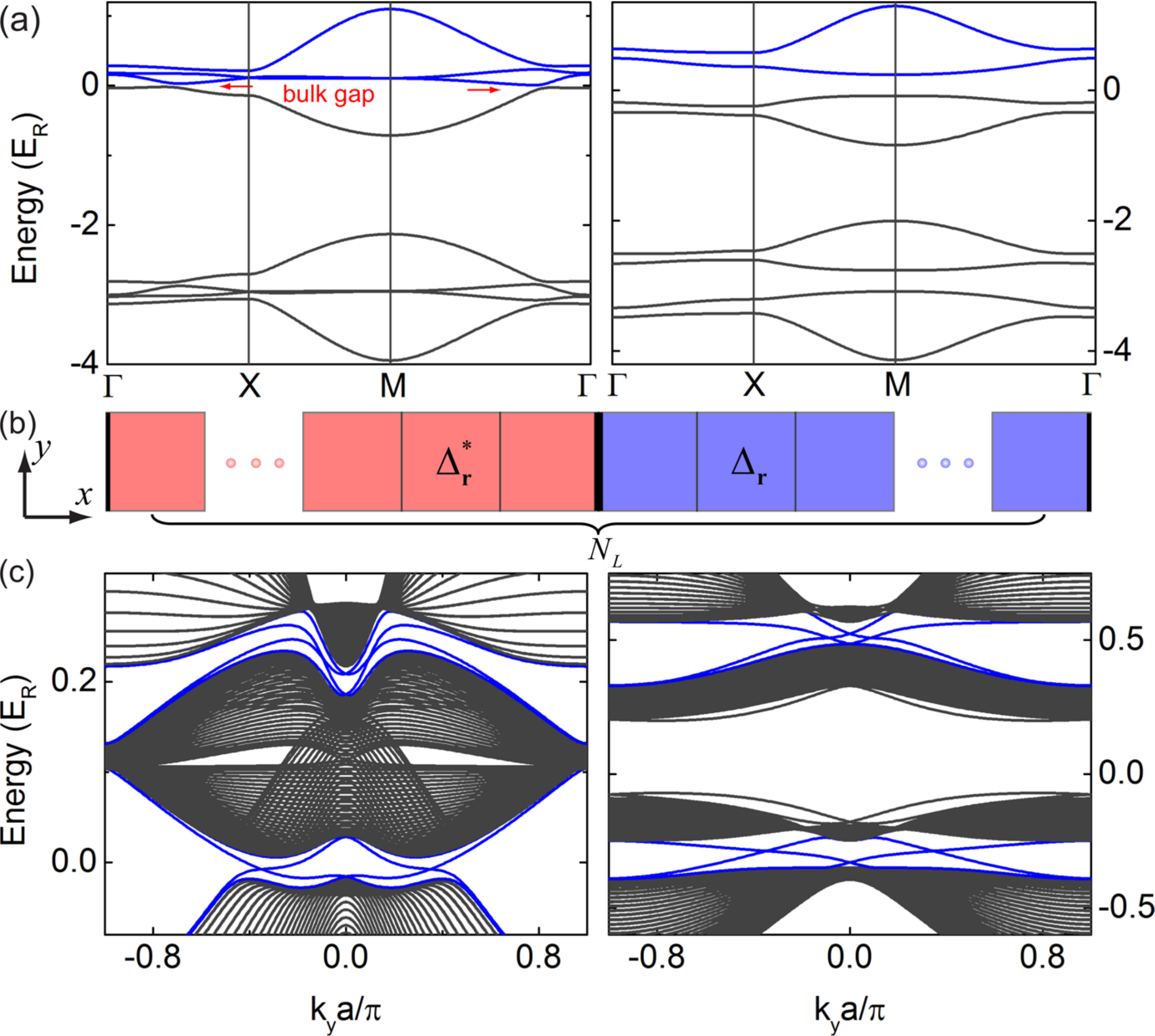}
\caption{ (color online). (a) Quasiparticle excitation spectra for two different odd-parity superfluids: $\rm TC_4S$ (left) and $\rm C_4S$ (right) along the high-symmetry lines in the first Brillouin zone, with $\mu_{\uparrow}=-1.5497\, E_R$, $\mu_{\downarrow}=-4.3957\, E_R$ and respectively $U=0.225$ and $0.315\, E_Ra^2$. A enlarged unit cell of a system in the presence of periodic domain walls, which exist in the center and the edges of the enlarged unit cell. Each one contains $N_L$ unit cells of the optical lattice, which is denoted by solid squares. The pairing order parameters on the left and right parts are time-reversal of each other. (c) Excitation spectra for the system in the presence of domain walls shown in (b) with same parameters used in (a). Blue solid lines denote the topological protected excitations at the domain walls. Here, we choose $k_x=0$ and $N_L=80$.}
\label{fig3}
\end{figure}

The band structure in Fig.~\ref{fig1} shows that the single-particle Bloch states at TRI points in between two Fermi surfaces are not degenerate. Two of them at $\Gamma$ and M points must be even-parity states, because the little groups at $\Gamma$ and M points coincide with the $D_4$ point group and parity-odd state should be two-fold degenerate~\cite{Sun2012}. Due to the $D_4$ symmetry of the lattice, the other two single-particle states at X and $\rm X'$ points should have the same eigenvalue of the parity operator. These lead to $\prod_{\alpha,n}\bar{p}_n(\Gamma_{\alpha})=1$. We also see that the spin-down Fermi surface encloses only one TRI point. In this sense, we identify that $\nu=1$ is  for the fully-gapped odd-parity superfluid phases, $\rm TC_4S$ and $\rm TC_2S$. That in turn indicates that  the two phases are topologically nontrivial. While for the fully-gapped $\rm C_4S$ phase, strong interaction induces a larger pairing order parameter, which changes the structure of excitations leading to different topology as shown in Fig.~\ref{fig3}. To prove this conclusion, we directly map out the topological edge excitations by artificially putting periodic domain walls in the system. Figure~\ref{fig3}(c) confirms our arguments that the $\rm TC_4S$ phase is topologically nontrivial. We further confirm that the $\rm TC_2S$ phase shows similar quasiparticle excitations and is also topologically nontrivial.

We would like to stress that for the weak coupling limit which applies to the topological phases $\rm TC_4S$ and $\rm TC_2S$~\cite{suppl}, the mean-field BCS theory should be valid and reliable even for a 2D system we considered. Otherwise, for the strong coupling limit, it is expected to be qualitatively correct based on what is widely known in the study of BCS-BEC crossover. Also, the strong interaction can undermine the assumption that the order parameter takes the same periodicity as the lattice. As detailed in~\cite{suppl}, the $\rm C_4S$ phase will be replaced by an even parity pairing phase.

{\it Experimental realization and detection.}---
To generate the desired lattice potential of Eq.~(\ref{potential}) in experiments, we merely need to provide a single blue-detuned linearly polarized monochromatic light beam, as described in Ref.~\cite{Sun2012}. The requirement that $V_2/V_1=1$ can be readily fulfilled. The maximum of the full pairing gap for the topological phases shown in Fig.~\ref{fig2} is about $0.01\, E_R$, which  corresponds to an experimentally feasible temperature scale of 10 nK. The odd-parity superfluids are characterized by the existence of edge states in domain walls or in the edges of a finite system confined in a box trap~\cite{Mukherjee2016}. By applying spatially resolved radio-frequency spectroscopy~\cite{Shin2007}, the signature of the edge states can be inferred from the local density of states~\cite{Liu2014}.

{\it Conclusion.}---We study fermion pairing for a spin imbalanced atomic Fermi gas loaded into a $D_4$ symmetric spin-independent bond-centered square optical lattice. Topological odd-parity superfluid phases spontaneously emerge from purely $s$-wave attractive interactions, in notable contrast to the conventional mechanism of topological superfluidity relying on interaction of high partial waves. Strong $s$-wave interaction can now be routinely realized in cold atomic gases via Feshbach resonances. The key ingredients for the topological superfluid phases presented here are: 1.~the existence of well matched Fermi surfaces crossing two neighboring energy bands and 2.~even and odd parities of the fermions close to the spin-up and down Fermi surfaces, respectively. These necessary prerequisites can be provided in an experimentally easily realizable square optical lattice. Our proposal prevents experimental complexities of previously discussed schemes of topological superfluidity, for example, the necessity of higher-partial-wave interactions, synthetic gauge fields and spin-dependent lattices.

This work is supported by NSFC (No.~11574100) and the National Thousand-Young-Talents Program (Z.-F.X.), and U.S. ARO (W911NF-11-1-0230), AFOSR (FA9550-16-1-0006), Overseas Collaboration Program
of NSF of China (No.~11429402) sponsored by Peking
University, and National Basic Research Program of China
(No.~2012CB922101) (W. V. L.).  A. H. acknowledges support
by DFG-SFB925 and the Hamburg Centre for Ultrafast
Imaging (CUI).


%

\clearpage
\onecolumngrid
\renewcommand\thefigure{S\arabic{figure}}
\setcounter{figure}{0}

{
\center \bf \large
Supplemental Material\\
\vspace*{0.25cm}
}

\newcommand{\mysetlabel}[2]{ \newcounter{maintextfig} \setcounter{maintextfig}{#2} \addtocounter{maintextfig}{-1}
\refstepcounter{maintextfig} \label{#1}}

In this supplementary material, we provide additional
details on (A) diagonalization of the mean-field Hamiltonian, (B) tight-binding models, (C) superfluid phases calculated from the tight-binding models.

\vspace{0.1in}
\centerline{\bf (A) Diagonalization of the mean-field Hamiltonian}
\vspace{0.1in}

In the main text, within the mean-field framework we obtain a mean-field Hamiltonian
\begin{eqnarray}
 \hat{H}_{\rm MF}
 =\sum\limits_{n\mathbf{k}\sigma}(\xi_{n\mathbf{k}}-\mu_{\sigma}) \hat{\psi}_{n\mathbf{k}\sigma}^{\dag}\hat{\psi}_{n\mathbf{k}\sigma}
 +\frac{\mathcal{V}}{U}\sum\limits_{\mathbf{Q}}|\Delta_{\mathbf{Q}}|^2
 +\sum\limits_{nm\mathbf{k}}\Big(\hat{\psi}_{n\mathbf{k}\uparrow}^{\dag} \hat{\psi}_{m,-\mathbf{k}\downarrow}^{\dag}\Delta_{nm\mathbf{k}}+\rm h.c.\Big).
\end{eqnarray}
It can be rewritten as
\begin{eqnarray}
 \hat{H}_{\rm MF}
 =\sum\limits_{\mathbf{k}}\hat{\Psi}_{\mathbf{k}}^{\dagger} \mathcal{H}_{\rm BdG}(\mathbf{k})\hat{\Psi}_{\mathbf{k}}+\sum_{n\mathbf{k}}(\xi_{n\mathbf{k}}-\mu_{\downarrow})
 +\frac{\mathcal{V}}{U}\sum_{\mathbf{Q}}|\Delta_{\mathbf{Q}}|^2,
\end{eqnarray}
where $\hat{\Psi}_{\mathbf{k}}=(\hat{\psi}_{1\mathbf{k}\uparrow}, \hat{\psi}_{2\mathbf{k}\uparrow},...,\hat{\psi}^{\dagger}_{1,-\mathbf{k}\downarrow}, \hat{\psi}^{\dagger}_{2,-\mathbf{k}\downarrow},...)^T$ and
\begin{eqnarray}
 \mathcal{H}_{\rm BdG}(\mathbf{k})
 =\left(
 \begin{array}{ccc|ccc}
 \xi_{1\mathbf{k}}-\mu_{\uparrow} &  &  & \Delta_{11\mathbf{k}} & \Delta_{12\mathbf{k}} & ...  \\
 & \xi_{2\mathbf{k}}-\mu_{\uparrow} &  & \Delta_{21\mathbf{k}} & \Delta_{22\mathbf{k}} & ...  \\
 & & \ddots & \vdots & \vdots & \ddots  \\
 \hline
 \Delta^*_{11\mathbf{k}} & \Delta^*_{21\mathbf{k}}  & ... & -\xi_{1\mathbf{k}}+\mu_{\downarrow} & &  \\
  \Delta^*_{12\mathbf{k}} & \Delta^*_{22\mathbf{k}}  & ... & & -\xi_{2\mathbf{k}}+\mu_{\downarrow} & \\
  \vdots & \vdots  & \ddots & & & \ddots\\
 \end{array}
 \right).
\end{eqnarray}
Diagonalizing $\mathcal{H}_{\rm BdG}(\mathbf{k})$, we obtain the quasiparticle excitation energy $\mathcal{E}_{n}(\mathbf{k})$.
We thus obtain the grand potential as
$\Omega=\sum_{n\mathbf{k}}(\xi_{n\mathbf{k}}-\mu_{\downarrow})
 +\frac{\mathcal{V}}{U}\sum_{\mathbf{Q}}|\Delta_{\mathbf{Q}}|^2 -\frac{1}{\beta}\sum_{n}\ln[1+\exp[-\beta \mathcal{E}_{n\mathbf{k}})]$.
Minimizing $\Omega$ with the simulated annealing method, we obtain the ground state. Numerically, we also use the self-consistent iteration method.
Initially, we randomly assume the value of the order parameter, later we recalculate it from the eigenstates of $\mathcal{H}_{\rm BdG}(\mathbf{k})$. The ground-state order parameter is obtained from tenths of independent converged iterations.

\vspace{0.1in}
\centerline{\bf (B) Tight-binding models}
\vspace{0.1in}

\begin{figure}[H]
\centering
\includegraphics[width=0.5\linewidth]{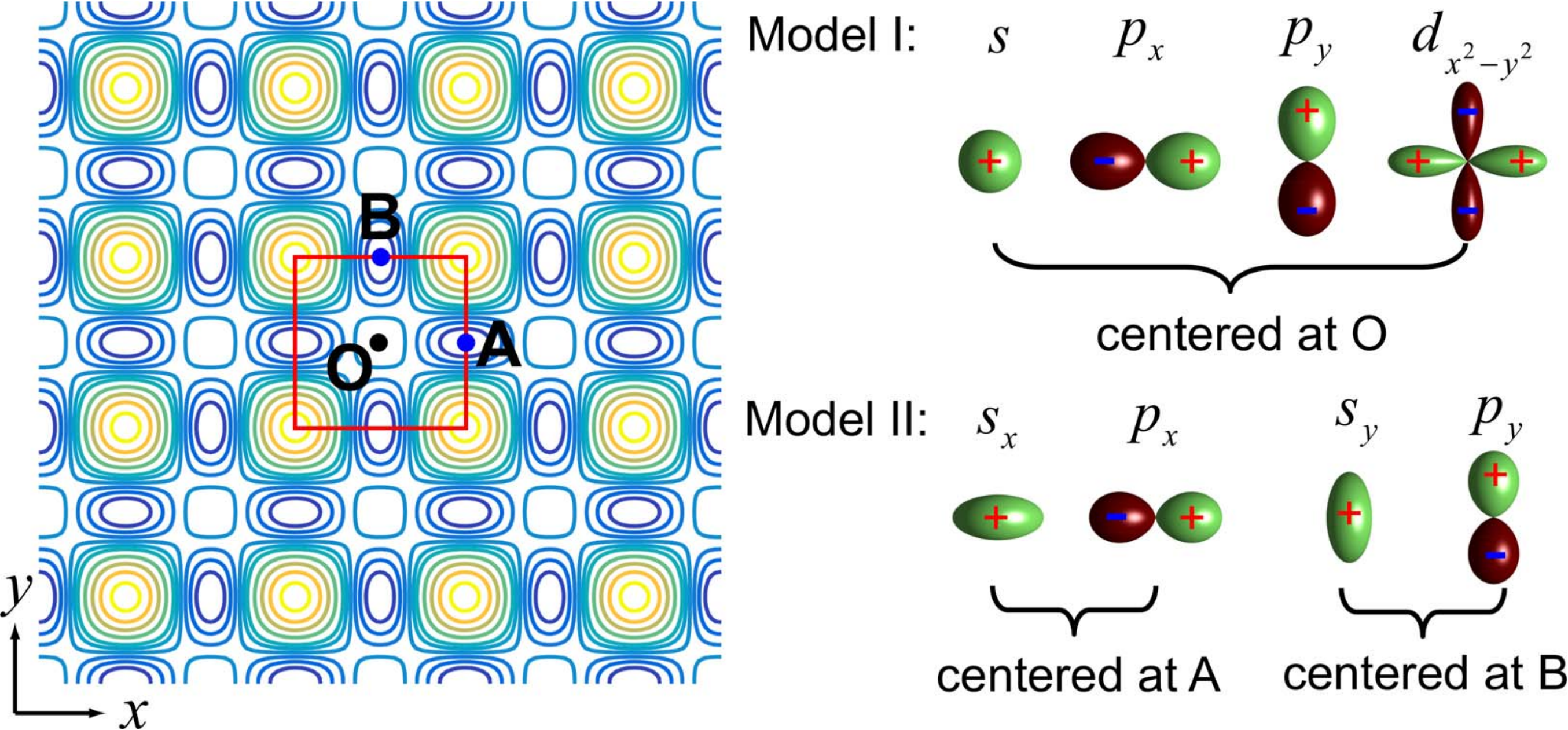}
\caption{(LEFT) Contour plot of the lattice potential. (Right) Wannier orbitals for two different tight-binding models for describing the lowest four energy bands. The first tight-binding model involves $s$, $p_x$, $p_y$, and $d_{x^2-y^2}$ orbitals whose centers are all located at O points. The second model involves $s_x$ and $p_x$ orbitals located at A points and $s_y$ and $p_y$ orbitals located at B points.}
\label{sfig1}
\end{figure}

To describe the lowest four energy bands of the lattice potential we considered, we take two different tight-binding models. For the model I, four orbitals ($s$, $p_x$, $p_y$, and $d_{x^2-y^2}$) centered at O points are involved. For the model II, we also consider four orbitals ($s_x$, $p_x$, $s_y$, and $p_y$). Two orbitals ($s_x$ and $p_x$) are centered at A points and the other two ($s_y$ and $p_y$) are centered at B points.

\begin{figure}[H]
\centering
\includegraphics[width=0.7\linewidth]{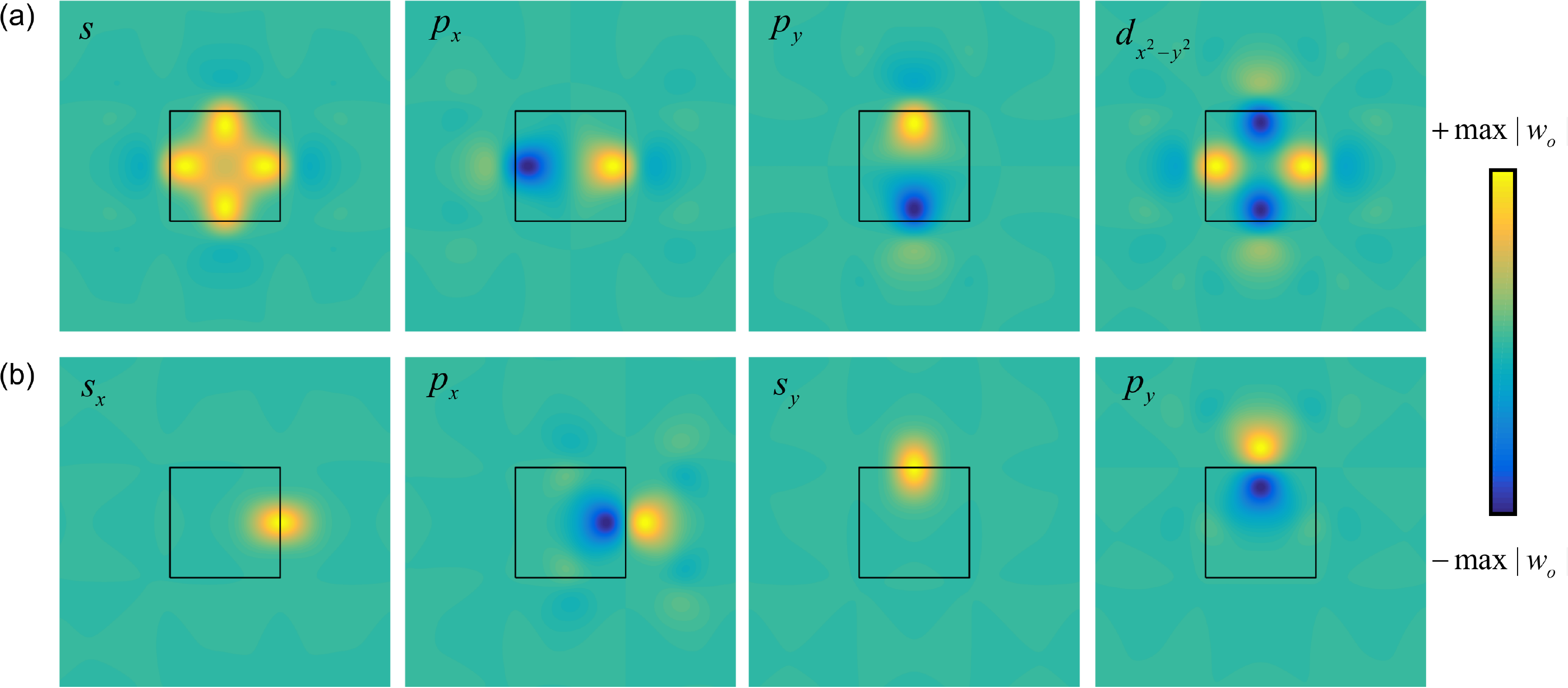}
\caption{Wannier functions $w_o$ for two different tight-binding models: I (a) and II (b). Here, $o=s, p_x, p_y, d_{x^2-y^2}$ for the model I and $o=s_x, p_x, s_y, p_y$ for the model II. Black solid square denotes the unit cell of the lattice centered at the O point.}
\label{sfig2}
\end{figure}

To obtain the tight-binding models, we first derive Wannier functions numerically. Here, our numerical method is based on Refs.~[42,43], where the Wannier functions are chosen as eigenstates of band-projected position operators $R_{1,2}$. We take $R_j=\mathcal{P}(\mathbf{b}_j\cdot \hat{\mathbf{r}})\mathcal{P}$, where $\hat{\mathbf{r}}=(\hat{x},\hat{y})^T$ are real-space position operators, $\mathbf{b}_j$ ($j=1,2$) are reciprocal lattice vectors, and $\mathcal{P}$ is the band projection operator. For the case we considered, the lowest-four bands are isolated from the higher bands. We thus choose the projection operator as $\mathcal{P}=\sum_{\alpha=1}^4\sum_{\mathbf{k}}|\mathbf{k},\alpha\rangle\langle\mathbf{k},\alpha|$, where $|\mathbf{k},\alpha\rangle$ is the Bloch state with quasimomentum $\mathbf{k}$ for the $\alpha$-th band. Different from that discussed in Ref.~\cite{Uehlinger2013}, by finding the simultaneous eigenstates of two position operators projected onto the lowest four bands, we still cannot fully determine four Wannier functions, because there are spatially overlapped Wannier functions, e.g. for the model I four Wannier functions are all centered at the O point. To solve this degeneracy, we further consider the point-group symmetry, i.e., mirror symmetry, of Wannier functions. For instance, by considering the operation $\mathcal{D}:(x,y)\rightarrow (-x,y)$, we can single out the $p_x$ orbital from $s$, $p_y$, and $d_{x^2-y^2}$ orbitals when they are the simultaneous eigenstates of $R_{1,2}$ with same eigenvalues. This is because only the $p_x$ orbital changes sign under the operation $\mathcal{D}$. The numerically calculated Wannier functions for the case with $V_1=V_2=5\,E_R$ are shown in Fig.~\ref{sfig2}. With Wannier functions, we obtain the tight-binding models where the tunneling between different lattice sites are calculated numerically.

Before showing the tight-binding models, we would like to emphasize that Wannier functions for two different models can be related with each other by a linear transformation:
\begin{eqnarray}
\left(\begin{array}{cccc}
w_{s}(\mathbf{r}-\mathbf{r}_O)\\
w_{p_x}(\mathbf{r}-\mathbf{r}_O)\\
w_{p_y}(\mathbf{r}-\mathbf{r}_O)\\
w_{d_{x^2-y^2}}(\mathbf{r}-\mathbf{r}_O)\\
\end{array}\right)=
\left(
\begin{array}{cccc}
\frac{1}{2} & \frac{1}{2} & \frac{1}{2} & \frac{1}{2} \\
-\frac{1}{\sqrt{2}} & 0 & 0 & \frac{1}{\sqrt{2}} \\
0 & \frac{1}{\sqrt{2}} & -\frac{1}{\sqrt{2}} & 0 \\
\frac{1}{2} & -\frac{1}{2} & -\frac{1}{2} & \frac{1}{2} \\
\end{array}
\right)
\times\frac{1}{\sqrt{2}}\left(
\begin{array}{cccc}
w_{s_x}(\mathbf{r}-\mathbf{r}_{\bar{A}})+w_{p_x}(\mathbf{r}-\mathbf{r}_{\bar{A}}) \\
w_{s_y}(\mathbf{r}-\mathbf{r}_{B})-w_{p_y}(\mathbf{r}-\mathbf{r}_{B}) \\
w_{s_y}(\mathbf{r}-\mathbf{r}_{\bar{B}})+w_{p_y}(\mathbf{r}-\mathbf{r}_{\bar{B}}) \\
w_{s_x}(\mathbf{r}-\mathbf{r}_{A})-w_{p_x}(\mathbf{r}-\mathbf{r}_{A}) \\
\end{array}
\right).
\end{eqnarray}
Here, $\mathbf{r}_O=(ma,na)$ ($m,n \in \mathbb{Z}$) is the position of the O points, $\mathbf{r}_A=\mathbf{r}_O+(a/2,0)$ and $\mathbf{r}_B=\mathbf{r}_O+(0,a/2)$ are positions of A and B points, and $\mathbf{r}_{\bar{A}}=\mathbf{r}_O-(a/2,0)$ and $\mathbf{r}_{\bar{B}}=\mathbf{r}_O-(0,a/2)$ denote another A and B points. $w_o(\mathbf{r}-\mathbf{r}_O)$ denotes the Wannier function centered at $\mathbf{r}_O$.

For the model I, in the momentum space the single-particle Hamiltonian takes the following form
\begin{eqnarray}
\hat{H}_0=\sum\limits_{\mathbf{k}}\left(
\begin{array}{cccc}
\hat{s}_{\mathbf{k}}^{\dagger} & \hat{p}^{\dagger}_{x,\mathbf{k}} & \hat{p}^{\dagger}_{y,\mathbf{k}} & \hat{d}^{\dagger}_{\mathbf{k}}
\end{array}\right)H_0(\mathbf{k})
\left(
\begin{array}{c}
\hat{s}_{\mathbf{k}}\\
\hat{p}_{x,\mathbf{k}}\\
\hat{p}_{y,\mathbf{k}}\\
\hat{d}_{\mathbf{k}}
\end{array}\right),
\end{eqnarray}
where $\hat{s}_{\mathbf{k}}$, $\hat{p}_{x,\mathbf{k}}$, $\hat{p}_{y,\mathbf{k}}$, and $\hat{d}_{\mathbf{k}}$ are the annihilation operators for the $s$, $p_x$, $p_y$ and $d_{x^2-y^2}$ orbitals at $\mathbf{k}$, respectively, and the matrix elements for the Hermitian matrix $H_0(\mathbf{k})$ are given by
\begin{eqnarray}
\ [H_0(\mathbf{k})]_{11}&=&\delta_s+2t_{ssx}[\cos(k_xa)+\cos(k_ya)] +2t_{ssxy}[\cos(k_xa+k_ya)+\cos(k_xa-k_ya)]\nonumber\\
&&+2t_{ss2x}[\cos(2k_xa)+\cos(2k_ya)]+2t_{ss2x2y}[\cos(2k_xa+2k_ya)+\cos(2k_xa-2k_ya)]\nonumber\\
&&+2t_{ss2xy}[\cos(2k_xa+k_ya)+\cos(2k_xa-k_ya)+\cos(k_xa+2k_ya)+\cos(k_xa-2k_ya)],\nonumber\\
\ [H_0(\mathbf{k})]_{22}&=&\delta_p+2t_{ppx}\cos(k_xa)+2t_{ppy}\cos(k_ya)+2t_{ppxy}[\cos(k_xa+k_ya)+\cos(k_xa-k_ya)]\nonumber\\
&&+2t_{pp2x}\cos(2k_xa)+2t_{pp2y}\cos(2k_ya)+2t_{pp2xy}[\cos(2k_xa+k_ya)+\cos(2k_xa-k_ya)]\nonumber\\
	&&+2t_{ppx2y}[\cos(k_xa+2k_ya)+\cos(k_xa-2k_ya)]+2t_{pp2x2y}[\cos(2k_xa+2k_ya)+\cos(2k_xa-2k_ya)],\nonumber\\
\ [H_0(\mathbf{k})]_{33}&=&\delta_p+2t_{ppx}\cos(k_ya)+2t_{ppy}\cos(k_xa) +2t_{ppxy}[\cos(k_xa+k_ya)+\cos(k_xa-k_ya)]\nonumber\\
&&+2t_{pp2x}\cos(2k_ya)+2t_{pp2y}\cos(2k_xa)+2t_{pp2xy}[\cos(k_xa+2k_ya)+\cos(k_xa-2k_ya)]\nonumber\\
&&+2t_{ppx2y}[\cos(2k_xa+k_ya)+\cos(2k_xa-k_ya)]+2t_{pp2x2y}[\cos(2k_xa+2k_ya)+\cos(2k_xa-2k_ya)],\nonumber\\
\ [H_0(\mathbf{k})]_{44}&=&\delta_d+2t_{ddx}[\cos(k_xa)+\cos(k_ya)] +2t_{ddxy}[\cos(k_xa+k_ya)+\cos(k_xa-k_ya)]\nonumber\\
&&+2t_{dd2x}[\cos(2k_xa)+\cos(2k_ya)]+2t_{dd2x2y}[\cos(2k_xa+2k_ya)+\cos(2k_xa-2k_ya)]\nonumber\\
&&+2t_{dd2xy}[\cos(2k_xa+k_ya)+\cos(2k_xa-k_ya)+\cos(k_xa+2*k_ya)+\cos(k_xa-2*k_ya)],\nonumber\\
\ [H_0(\mathbf{k})]_{12}&=&2it_{spx}\sin(k_xa)+2it_{spxy}[\sin(k_xa+k_ya)+\sin(k_xa-k_ya)]\nonumber\\
&&+2it_{sp2x}\sin(2k_xa)+2it_{sp2xy}[\sin(2k_xa+k_ya)+\sin(2k_xa-k_ya)]\nonumber\\
&&+2it_{spx2y}[\sin(k_xa+2k_ya)+\sin(k_xa-2k_ya)]+2it_{sp2x2y}[\sin(2k_xa+2k_ya)+\sin(2k_xa-2k_ya)],\nonumber\\
\ [H_0(\mathbf{k})]_{13}&=&2it_{spx}\sin(k_ya)+2it_{spxy}[\sin(k_xa+k_ya)-\sin(k_xa-k_ya)]\nonumber\\
&&+2it_{sp2x}\sin(2k_ya)+2it_{sp2xy}[\sin(k_xa+2k_ya)-\sin(k_xa-2k_ya)]\nonumber\\
&&+2it_{spx2y}[\sin(2k_xa+k_ya)-\sin(2k_xa-k_ya)]+2it_{sp2x2y}[\sin(2k_xa+2k_ya)-\sin(2k_xa-2k_ya)],\nonumber\\
\ [H_0(\mathbf{k})]_{14}&=&2t_{sdx}[\cos(k_xa)-\cos(k_ya)]+2t_{sd2x}[\cos(2k_xa)-\cos(2k_ya)]\nonumber\\
&&+2t_{sd2xy}[\cos(2k_xa+k_ya)+\cos(2k_xa-k_ya)-\cos(k_xa+2k_ya)-\cos(k_xa-2k_ya)],\nonumber\\
\ [H_0(\mathbf{k})]_{23}&=&2t_{pxpyxy}[\cos(k_xa+k_ya)-\cos(k_xa-k_ya)]+2t_{pxpy2xy}[\cos(2k_xa+k_ya)-\cos(2k_xa-k_ya)]\nonumber\\
&&+2t_{pxpyx2y}[\cos(k_xa+2k_ya)-\cos(k_xa-2k_ya)]+2t_{pxpy2x2y}[\cos(2k_xa+2k_ya)-\cos(2k_xa-2k_ya)],\nonumber\\
\ [H_0(\mathbf{k})]_{24}&=&2it_{pdx}\sin(k_xa)+2it_{pdxy}[\sin(k_xa+k_ya)+\sin(k_xa-k_ya)]\nonumber\\
&&+2it_{pd2x}\sin(2k_xa)+2it_{pd2xy}[\sin(2k_xa+k_ya)+\sin(2k_xa-k_ya)]\nonumber\\
&&+2it_{pdx2y}[\sin(k_xa+2k_ya)+\sin(k_xa-2k_ya)]+2it_{pd2x2y}[\sin(2k_xa+2k_ya)+\sin(2k_xa-2k_ya)],\nonumber\\
\ [H_0(\mathbf{k})]_{34}&=&-2it_{pdx}\sin(k_ya)-2it_{pdxy}[\sin(k_xa+k_ya)-\sin(k_xa-k_ya)]\nonumber\\
&&-2it_{pd2x}\sin(2k_ya)-2it_{pd2xy}[\sin(k_xa+2k_ya)-\sin(k_xa-2k_ya)]\nonumber\\
&&-2it_{pdx2y}[\sin(2k_xa+k_ya)-\sin(2k_xa-k_ya)]-2it_{pd2x2y}[\sin(2k_xa+2k_ya)-\sin(2k_xa-2k_ya)],\nonumber\\
\ [H_0(\mathbf{k})]_{21}&=&\left([H_0(\mathbf{k})]_{12}\right)^*,\nonumber\\
\ [H_0(\mathbf{k})]_{31}&=&\left([H_0(\mathbf{k})]_{13}\right)^*,\nonumber\\
\ [H_0(\mathbf{k})]_{32}&=&\left([H_0(\mathbf{k})]_{23}\right)^*,\nonumber\\
\ [H_0(\mathbf{k})]_{41}&=&\left([H_0(\mathbf{k})]_{14}\right)^*,\nonumber\\
\ [H_0(\mathbf{k})]_{42}&=&\left([H_0(\mathbf{k})]_{24}\right)^*,\nonumber\\
\ [H_0(\mathbf{k})]_{43}&=&\left([H_0(\mathbf{k})]_{34}\right)^*.
\end{eqnarray}
Here, $\delta_s$, $\delta_p$, and $\delta_d$ represent the on-site energies for $s$, $p$, and $d$ orbitals, respectively.

For the model II, in the momentum space the single-particle Hamiltonian takes the following form
\begin{eqnarray}
\hat{H}_0=\sum\limits_{\mathbf{k}}\left(
\begin{array}{cccc}
\hat{s}_{x,\mathbf{k}}^{\dagger} & \hat{s}^{\dagger}_{y,\mathbf{k}} & \hat{p}^{\dagger}_{x,\mathbf{k}} & \hat{p}^{\dagger}_{y,\mathbf{k}}
\end{array}\right)H_0(\mathbf{k})
\left(
\begin{array}{c}
\hat{s}_{x,\mathbf{k}}\\
\hat{s}_{y,\mathbf{k}}\\
\hat{p}_{x,\mathbf{k}}\\
\hat{p}_{y,\mathbf{k}}
\end{array}\right),
\end{eqnarray}
where $\hat{s}_{x,\mathbf{k}}$, $\hat{s}_{y,\mathbf{k}}$, $\hat{p}_{x,\mathbf{k}}$, and $\hat{p}_{y,\mathbf{k}}$ are the annihilation operators for the $s_x$, $s_y$, $p_x$ and $p_{y}$ orbitals at $\mathbf{k}$, respectively, and the matrix elements for the Hermitian matrix $H_0(\mathbf{k})$ are given by
\begin{eqnarray}
\ [H_0(\mathbf{k})]_{11}&=&\delta_s-2t_{ss1}\cos(k_xa)-2t_{ss3}\cos(k_ya)-2t_{ss4}[\cos(k_xa+k_ya)+\cos(k_xa-k_ya)],\nonumber\\
\ [H_0(\mathbf{k})]_{22}&=&\delta_s-2t_{ss3}\cos(k_xa)-2t_{ss1}\cos(k_ya)-2t_{ss4}[\cos(k_xa+k_ya)+\cos(k_xa-k_ya)],\nonumber\\
\ [H_0(\mathbf{k})]_{33}&=&\delta_p+2t_{pp1}\cos(k_xa)-2t_{pp3}\cos(k_ya)+2t_{pp4}[\cos(k_xa+k_ya)+\cos(k_xa-k_ya)],\nonumber\\
\ [H_0(\mathbf{k})]_{44}&=&\delta_p-2t_{pp3}\cos(k_xa)+2t_{pp1}\cos(k_ya)+2t_{pp4}[\cos(k_xa+k_ya)+\cos(k_xa-k_ya)],\nonumber\\
\ [H_0(\mathbf{k})]_{12}&=&-2t_{ss2}[\cos(k_xa/2+k_ya/2)+\cos(k_xa/2-k_ya/2)],\nonumber\\
\ [H_0(\mathbf{k})]_{13}&=&2it_{sp1}\sin(k_xa)+2it_{sp4}[\sin(k_xa+k_ya)+\sin(k_xa-k_ya)],\nonumber\\
\ [H_0(\mathbf{k})]_{14}&=&2it_{sp2}[\sin(k_xa/2+k_ya/2)-\sin(k_xa/2-k_ya/2)],\nonumber\\
\ [H_0(\mathbf{k})]_{23}&=&2it_{sp2}[\sin(k_xa/2+k_ya/2)+\sin(k_xa/2-k_ya/2)],\nonumber\\
\ [H_0(\mathbf{k})]_{24}&=&2it_{sp1}\sin(k_ya)+2it_{sp4}[\sin(k_xa+k_ya)-\sin(k_xa-k_ya)],\nonumber\\
\ [H_0(\mathbf{k})]_{34}&=&2t_{pp2}[\cos(k_xa/2+k_ya/2)-\cos(k_xa/2-k_ya/2)],\nonumber\\
\ [H_0(\mathbf{k})]_{21}&=&\left([H_0(\mathbf{k})]_{12}\right)^*,\nonumber\\
\ [H_0(\mathbf{k})]_{31}&=&\left([H_0(\mathbf{k})]_{13}\right)^*,\nonumber\\
\ [H_0(\mathbf{k})]_{32}&=&\left([H_0(\mathbf{k})]_{23}\right)^*,\nonumber\\
\ [H_0(\mathbf{k})]_{41}&=&\left([H_0(\mathbf{k})]_{14}\right)^*,\nonumber\\
\ [H_0(\mathbf{k})]_{42}&=&\left([H_0(\mathbf{k})]_{24}\right)^*,\nonumber\\
\ [H_0(\mathbf{k})]_{43}&=&\left([H_0(\mathbf{k})]_{34}\right)^*.
\end{eqnarray}
Here, $\delta_s$ and $\delta_p$ represent the on-site energies for $s$ and $p$ orbitals, respectively.

\begin{figure}[H]
\centering
\includegraphics[width=0.5\linewidth]{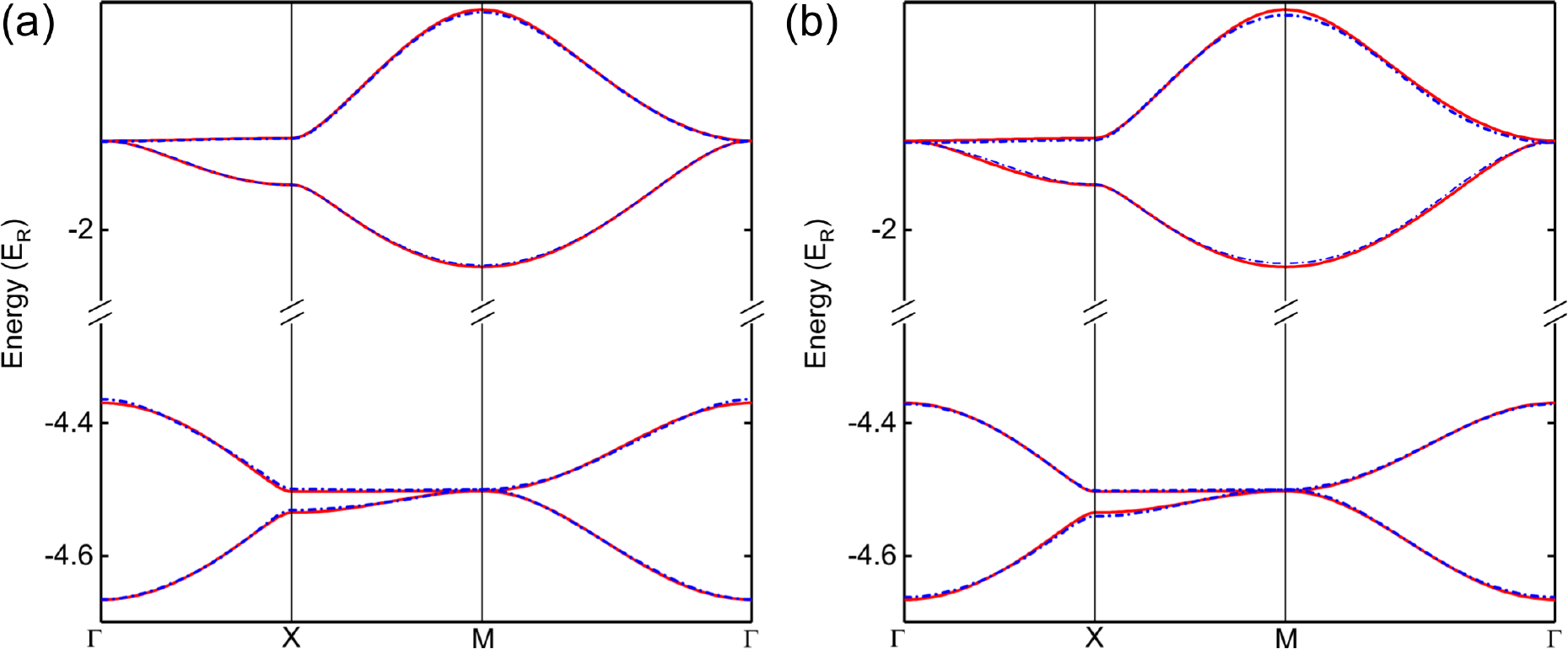}
\caption{Comparison between the exact band structure (red solid lines) obtained from plane-wave expansion and that calculated from the tight-binding models (blue dash-dotted lines) : I (a) and II (b) for $V_1=V_2=5\, E_R$.}
\label{sfig3}
\end{figure}

Numerically, the on-site energies and hopping coefficients are all calculated based on the obtained Wannier functions. To well describe the lowest four bands, we consider upto the farmost hopping from $(x,y)$ to $(x\pm 2a,y\pm 2a)$ for the model I. While for the model II, we consider only upto to next-next-nearest neighbor hopping: $(x,y)\rightarrow (x\pm a,y\pm a)$. Figure~\ref{sfig3} shows our numerical results. We confirm that both models can well reproduce the lowest four energy bands for the case of $V_1=V_2=5\, E_R$.

For general cases, as long as lowest four bands are isolated from other bands, tight-binding models involves four orbitals should well describe the system. For a small ratio $V_2/V_1$, the lattice potential turns out to be the conventional square lattice. Therefore, model I is more better to describe the system. While for a larger ratio, e.g., $V_2/V_1=1$ we considered, the Wannier functions for the model II are more spatially localized as illustrated in Fig.~\ref{sfig2}. We thus can well describe the band structure with the model II by including a fewer hopping terms.

\vspace{0.1in}
\centerline{\bf (C) Superfluid phases}
\vspace{0.1in}

In this section, we apply the tight-binding models to obtain the ground-state phases. For the model I, the field operator can be expanded by Wannier functions as
\begin{eqnarray}
\hat{\psi}_{\mathbf{r}\sigma}=\sum\limits_{\mathbf{r}_O}
w_s(\mathbf{r}-\mathbf{r}_O)\hat{s}_{\mathbf{r}_O,\sigma} +w_{p_x}(\mathbf{r}-\mathbf{r}_O)\hat{p}_{x,\mathbf{r}_O,\sigma}
+w_{p_y}(\mathbf{r}-\mathbf{r}_O)\hat{p}_{y,\mathbf{r}_O,\sigma}
+w_{d_{x^2-y^2}}(\mathbf{r}-\mathbf{r}_O)\hat{d}_{\mathbf{r}_O,\sigma},
\end{eqnarray}
where $\hat{s}_{\mathbf{r}_O,\sigma}$, $\hat{p}_{x,\mathbf{r}_O,\sigma}$, $\hat{p}_{y,\mathbf{r}_O,\sigma}$, and $\hat{d}_{\mathbf{r}_O,\sigma}$ represent the annihilation operators for fermions at orbitals $s$, $p_x$, $p_y$, and $d_{x^2-y^2}$ centered at $\mathbf{r}_O$.
Consider only the on-site interaction, we obtain the interaction-part Hamiltonian
\begin{eqnarray}
\hat{H}_{\rm int}&=&
-U\sum\limits_{\mathbf{r}_O}\int d\mathbf{r}\left[w_s(\mathbf{r}-\mathbf{r}_O)\hat{s}_{\mathbf{r}_O,\uparrow}^{\dagger} +w_{p_x}(\mathbf{r}-\mathbf{r}_O)\hat{p}_{x,\mathbf{r}_O,\uparrow}^{\dagger}
+w_{p_y}(\mathbf{r}-\mathbf{r}_O)\hat{p}_{y,\mathbf{r}_O,\uparrow}^{\dagger}
+w_{d_{x^2-y^2}}(\mathbf{r}-\mathbf{r}_O)\hat{d}_{\mathbf{r}_O,\uparrow}^{\dagger}
\right]\nonumber\\
&&\qquad\qquad\quad\times
\left[w_s(\mathbf{r}-\mathbf{r}_O)\hat{s}_{\mathbf{r}_O,\downarrow}^{\dagger} +w_{p_x}(\mathbf{r}-\mathbf{r}_O)\hat{p}_{x,\mathbf{r}_O,\downarrow}^{\dagger}
+w_{p_y}(\mathbf{r}-\mathbf{r}_O)\hat{p}_{y,\mathbf{r}_O,\downarrow}^{\dagger}
+w_{d_{x^2-y^2}}(\mathbf{r}-\mathbf{r}_O)\hat{d}_{\mathbf{r}_O,\downarrow}^{\dagger}\right]
\nonumber\\
&&\qquad\qquad\quad\times
\left[w_s(\mathbf{r}-\mathbf{r}_O)\hat{s}_{\mathbf{r}_O,\downarrow} +w_{p_x}(\mathbf{r}-\mathbf{r}_O)\hat{p}_{x,\mathbf{r}_O,\downarrow}
+w_{p_y}(\mathbf{r}-\mathbf{r}_O)\hat{p}_{y,\mathbf{r}_O,\downarrow}
+w_{d_{x^2-y^2}}(\mathbf{r}-\mathbf{r}_O)\hat{d}_{\mathbf{r}_O,\downarrow}\right]
\nonumber\\
&&\qquad\qquad\quad\times
\left[w_s(\mathbf{r}-\mathbf{r}_O)\hat{s}_{\mathbf{r}_O,\uparrow} +w_{p_x}(\mathbf{r}-\mathbf{r}_O)\hat{p}_{x,\mathbf{r}_O,\uparrow}
+w_{p_y}(\mathbf{r}-\mathbf{r}_O)\hat{p}_{y,\mathbf{r}_O,\uparrow}
+w_{d_{x^2-y^2}}(\mathbf{r}-\mathbf{r}_O)\hat{d}_{\mathbf{r}_O,\uparrow}
\right].
\end{eqnarray}
After expansion, we obtain 256 terms. Numerically, we keep all terms with nonzero coefficients. As discussed in our manuscript, the spin-up and spin-down chemical potentials are adjusted simultaneously to match the enclosed area of the two Fermi surfaces in momentum space. Therefore, the possibility of FFLO states is suppressed. We then consider the conventional BCS pairing and assume that the order parameter takes the same periodicity as the optical lattice potential.

\begin{figure}[t]
\centering
\includegraphics[width=0.7\linewidth]{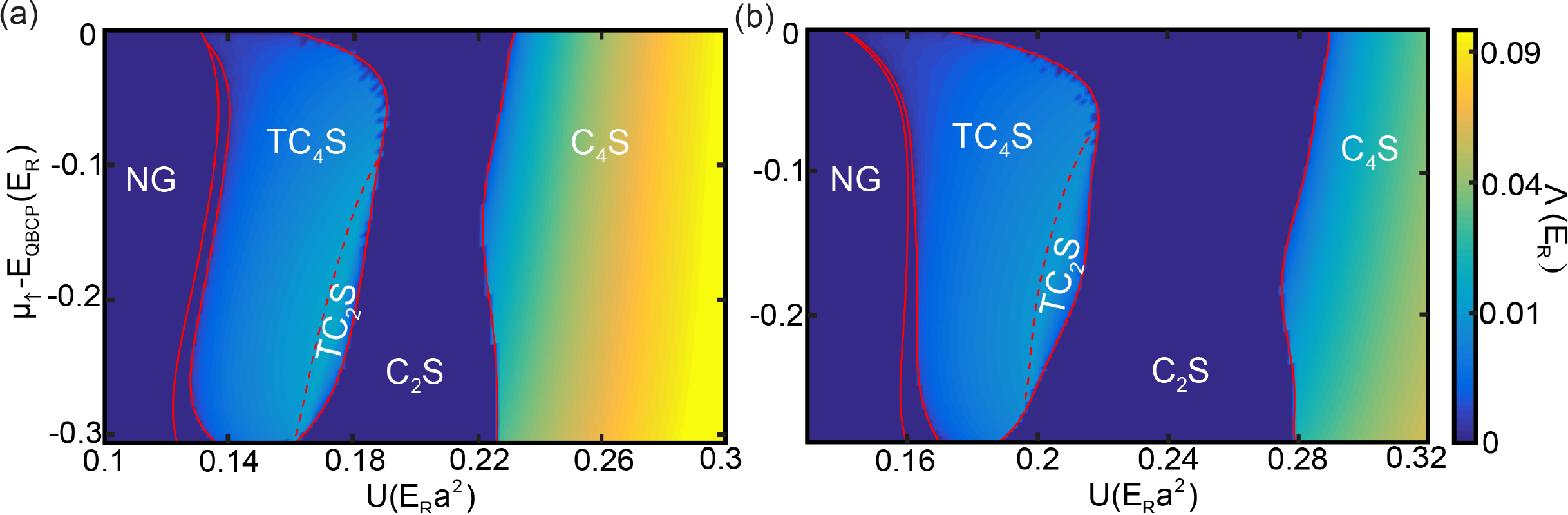}
\caption{Zero-temperature ground-state phase diagram by varying the s-wave interaction and the chemical potentials with tight-binding models I (a) and II (b).}
\label{sfig4}
\end{figure}

We further define the pairing order parameters as
\begin{eqnarray}
\Delta=\left(
\begin{array}{cccc}
-U_{\rm ss}\langle\hat{s}_{\downarrow}\hat{s}_{\uparrow}\rangle &
-U_{\rm sp}\langle\hat{p}_{x,\downarrow}\hat{s}_{\uparrow}\rangle &
-U_{\rm sp}\langle\hat{p}_{y,\downarrow}\hat{s}_{\uparrow}\rangle &
-U_{\rm sd}\langle\hat{d}_{\downarrow}\hat{s}_{\uparrow}\rangle \\
-U_{\rm sp}\langle\hat{s}_{\downarrow}\hat{p}_{x,\uparrow}\rangle &
-U_{\rm pp}\langle\hat{p}_{x,\downarrow}\hat{p}_{x,\uparrow}\rangle &
-U'_{\rm pp}\langle\hat{p}_{y,\downarrow}\hat{p}_{x\uparrow}\rangle &
-U_{\rm pd}\langle\hat{d}_{\downarrow}\hat{p}_{x,\uparrow}\rangle \\
-U_{\rm sp}\langle\hat{s}_{\downarrow}\hat{p}_{y,\uparrow}\rangle &
-U'_{\rm pp}\langle\hat{p}_{x,\downarrow}\hat{p}_{y,\uparrow}\rangle &
-U_{\rm pp}\langle\hat{p}_{y,\downarrow}\hat{p}_{y,\uparrow}\rangle &
-U_{\rm pd}\langle\hat{d}_{\downarrow}\hat{p}_{y,\uparrow}\rangle \\
-U_{\rm sd}\langle\hat{s}_{\downarrow}\hat{d}_{\uparrow}\rangle &
-U_{\rm pd}\langle\hat{p}_{x,\downarrow}\hat{d}_{\uparrow}\rangle &
-U_{\rm pd}\langle\hat{p}_{y,\downarrow}\hat{d}_{\uparrow}\rangle &
-U_{\rm dd}\langle\hat{d}_{\downarrow}\hat{d}_{\uparrow}\rangle
\end{array}
\right),
\end{eqnarray}
where we omit the subscript $\mathbf{r}_O$ because we are searching for the pairing order parameter showing the same periodicity as the lattice. The interaction parameters are derived based on the obtained Wannier functions. For instance, $U_{ss}=U\int d\mathbf{r} w_s^4(\mathbf{r})$, $U_{sp}=U\int d\mathbf{r} w_s^2(\mathbf{r})w_{p_x}^2(\mathbf{r})=U\int d\mathbf{r} w_s^2(\mathbf{r})w_{p_y}^2(\mathbf{r})$, and $U'_{pp}=U\int d\mathbf{r} w_{p_x}^2(\mathbf{r})w_{p_y}^2(\mathbf{r})$.

Applying the mean-field approximation, we obtain the ground states via self-consistent iteration method. The corresponding phase diagram is shown in Fig.~\ref{sfig4}(a). Similarly, five different phases: $\rm NG$, $\rm TC_4S$, $\rm TC_2S$, $\rm C_4S$, and $\rm C_2S$, are found. The structure of the phase diagram is similar to that shown in Fig.~2 of the manuscript, although quantitatively the phase boundaries are shifted slightly away from that derived from the plane-wave expansion. We attribute this deviation to the ignorance of the interaction among different lattice sites. For example, we choose $U=0.1657\, E_Ra^2$, $\mu_{\uparrow}=-1.475\,E_R$, and $\mu_{\downarrow}=4.383\,E_R$. The ground state falls into the topological superfluid phase $\rm TC_4S$. The corresponding order parameters are given by
\begin{eqnarray}
\Delta\simeq\left(
\begin{array}{cccc}
0 & 0.0044 & 0.0044i & 0 \\
0.0016 & 0 & 0 & 0.0094 \\
0.0016i & 0 & 0 & -0.0094i \\
0 & 0.0022 & -0.0022i & 0
\end{array}
\right)\, E_R,
\end{eqnarray}
We thus infer that for the topological superfluid phase $\rm TC_4S$, pairing is among even and odd-parity orbitals, and the phase difference between two odd orbitals is locked at $\pm \pi/2$, e.g. $\langle \hat{s}_{\downarrow}\hat{p}_{x,\uparrow}\rangle =-i\langle\hat{s}_{\downarrow}\hat{p}_{y,\uparrow}\rangle$
and $\langle \hat{d}_{\downarrow}\hat{p}_{x,\uparrow}\rangle =i\langle\hat{d}_{\downarrow}\hat{p}_{y,\uparrow}\rangle$.

For the model II, the field operator can also be expanded by Wannier functions as
\begin{eqnarray}
\hat{\psi}_{\mathbf{r}\sigma}=\sum\limits_{\mathbf{r}_A}\left[
w_{s_x}(\mathbf{r}-\mathbf{r}_A)\hat{s}_{\mathbf{r}_A,\sigma} +w_{p_x}(\mathbf{r}-\mathbf{r}_A)\hat{p}_{x,\mathbf{r}_A,\sigma}\right]
+\sum\limits_{\mathbf{r}_B}\left[w_{s_y}(\mathbf{r}-\mathbf{r}_B)\hat{s}_{y,\mathbf{r}_B,\sigma}
+w_{p_y}(\mathbf{r}-\mathbf{r}_B)\hat{p}_{y,\mathbf{r}_B,\sigma}\right],
\end{eqnarray}
where $\hat{s}_{x,\mathbf{r}_A,\sigma}$, $\hat{s}_{y,\mathbf{r}_B,\sigma}$, $\hat{p}_{x,\mathbf{r}_A,\sigma}$, and $\hat{p}_{y,\mathbf{r}_B,\sigma}$ represent the annihilation operators for fermions at orbitals $s_x$ and $p_x$ centered at $\mathbf{r}_A$ and $s_y$, and $p_y$ centered at $\mathbf{r}_B$.
Consider only the on-site interaction, we obtain the interaction-part Hamiltonian
\begin{eqnarray}
\hat{H}_{\rm int}&=&
-U\sum\limits_{\mathbf{r}_A}\int d\mathbf{r}\left[w_{s_x}(\mathbf{r}-\mathbf{r}_A)\hat{s}_{x,\mathbf{r}_A,\uparrow}^{\dagger} +w_{p_x}(\mathbf{r}-\mathbf{r}_A)\hat{p}_{x,\mathbf{r}_A,\uparrow}^{\dagger}
\right]
\left[w_{s_x}(\mathbf{r}-\mathbf{r}_A)\hat{s}_{x,\mathbf{r}_A,\downarrow}^{\dagger} +w_{p_x}(\mathbf{r}-\mathbf{r}_A)\hat{p}_{x,\mathbf{r}_A,\downarrow}^{\dagger}
\right]
\nonumber\\
&&\qquad\qquad\quad\times
\left[w_{s_x}(\mathbf{r}-\mathbf{r}_A)\hat{s}_{x,\mathbf{r}_A,\downarrow} +w_{p_x}(\mathbf{r}-\mathbf{r}_A)\hat{p}_{x,\mathbf{r}_A,\downarrow}
\right]
\left[w_{s_x}(\mathbf{r}-\mathbf{r}_A)\hat{s}_{x,\mathbf{r}_A,\uparrow} +w_{p_x}(\mathbf{r}-\mathbf{r}_A)\hat{p}_{x,\mathbf{r}_A,\uparrow}
\right]\nonumber\\
&&
-U\sum\limits_{\mathbf{r}_B}\int d\mathbf{r}\left[w_{s_y}(\mathbf{r}-\mathbf{r}_B)\hat{s}_{y,\mathbf{r}_B,\uparrow}^{\dagger} +w_{p_y}(\mathbf{r}-\mathbf{r}_B)\hat{p}_{y,\mathbf{r}_B,\uparrow}^{\dagger}
\right]
\left[w_{s_y}(\mathbf{r}-\mathbf{r}_B)\hat{s}_{y,\mathbf{r}_B,\downarrow}^{\dagger} +w_{p_y}(\mathbf{r}-\mathbf{r}_B)\hat{p}_{y,\mathbf{r}_B,\downarrow}^{\dagger}
\right]
\nonumber\\
&&\qquad\qquad\quad\times
\left[w_{s_y}(\mathbf{r}-\mathbf{r}_B)\hat{s}_{y,\mathbf{r}_B,\downarrow} +w_{p_y}(\mathbf{r}-\mathbf{r}_B)\hat{p}_{y,\mathbf{r}_B,\downarrow}
\right]
\left[w_{s_y}(\mathbf{r}-\mathbf{r}_B)\hat{s}_{y,\mathbf{r}_B,\uparrow} +w_{p_y}(\mathbf{r}-\mathbf{r}_B)\hat{p}_{y,\mathbf{r}_B,\uparrow}
\right].
\end{eqnarray}
After expansion, we keep all terms with nonzero coefficients.
We first search for the ground state with order parameters showing the same periodicity as the lattice. The pairing order parameters are defined as
\begin{eqnarray}
\Delta=\left(
\begin{array}{cccc}
-U_{\rm ss}\langle\hat{s}_{x,\downarrow}\hat{s}_{x,\uparrow}\rangle &
&
-U_{\rm sp}\langle\hat{p}_{x,\downarrow}\hat{s}_{x,\uparrow}\rangle &
\\
&
-U_{\rm ss}\langle\hat{s}_{y,\downarrow}\hat{s}_{y,\uparrow}\rangle &
&
-U_{\rm sp}\langle\hat{p}_{y,\downarrow}\hat{s}_{y,\uparrow}\rangle\\
-U_{\rm sp}\langle\hat{s}_{x,\downarrow}\hat{p}_{x,\uparrow}\rangle &
&
-U_{\rm pp}\langle\hat{p}_{x,\downarrow}\hat{p}_{x,\uparrow}\rangle &
\\
&
-U_{\rm sp}\langle\hat{s}_{y,\downarrow}\hat{p}_{y,\uparrow}\rangle &
&
-U_{\rm pp}\langle\hat{p}_{y,\downarrow}\hat{p}_{y,\uparrow}\rangle\\
\end{array}
\right),
\label{sop}
\end{eqnarray}
where we omit the subscript $\mathbf{r}_A$ and $\mathbf{r}_B$ for simplicity. The interaction parameters are derived from the obtained Wannier functions. For instance $U_{\rm ss}=U\int d\mathbf{r} w_{s_x}^4(\mathbf{r}-\mathbf{r}_A)$, $U_{\rm pp}=U\int d\mathbf{r} w_{p_x}^4(\mathbf{r}-\mathbf{r}_B)$, and $U_{\rm sp}=U\int d\mathbf{r} w_{s_x}^2(\mathbf{r}-\mathbf{r}_A)w_{p_x}^2(\mathbf{r}-\mathbf{r}_A)$. Again, we apply the mean-field approximation to obtain the ground state. Figure.~\ref{sfig4}(b) shows the phase diagram, which is similar to that shown in Fig.~2 of the manuscript where the plane-wave expansion method is applied. Quantitatively, the phase boundaries are more closer to that by using the plane-wave expansion method due to the more localized Wannier functions comparing to Wannier functions of the model I. For example, we choose $U=0.1974\, E_Ra^2$, $\mu_{\uparrow}=-1.4762\, E_R$, and $\mu_{\downarrow}=-4.3858\,E_R$. The ground state falls into the $\rm TC_4S$ phase. The corresponding pairing order parameters are
\begin{eqnarray}
\Delta\simeq\left(
\begin{array}{cccc}
0.0 &  & 0.0 &  \\
  & 0.0 &   & \quad 0.0  \\
0.0256 &   & \quad 0.0 &   \\
  & 0.0256i &   & \quad 0.0
\end{array}
\right)\, E_R,
\end{eqnarray}
We thus infer that for the $\rm TC_4S$ phase the pairing is among the spin-down fermions at $s$-orbtials and spin-up fermions at $p$-orbitals. The phase difference between two order parameters $\langle \hat{s}_{x,\downarrow}\hat{p}_{x,\uparrow}\rangle$ and $\langle \hat{s}_{y,\downarrow}\hat{p}_{y,\uparrow}\rangle$ are locked at $\pm \pi/2$ to satisfy the $C_4$ rotational symmetry of the ground state.

\vspace{0.1in}
{\bf Parity of the pairing order parameter.}---We now re-examine the parity of the pairing order parameter. The bond-centered square optical lattice preserves the $D_4$ point-group symmetry. Thus, there is a quadratic band crossing at $\Gamma$ point connecting the 3rd and the 4th band. With help of the tight-binding model I, we could infer that when the spin-up Fermi surface is tuned to be close to the quadratic band crossing, the spin-up fermions close to the Fermi surface are mainly composed by the odd-parity $p$ orbitals. In contrast, close to the spin-down Fermi surface which is tuned to lie near the maximum of the 2nd band, the fermions are mainly composed by even-parity orbitals. Here, the inversion center is chosen to be the $O$ point. In the weak coupling limit, the pairing is mainly among fermions close to the Fermi surfaces, and hence Cooper pair takes place mainly between odd-parity spin-up fermions and even-parity spin-down fermions, leading to odd-parity superfluidity.

Applying the tight-binding model II, we find that the 3rd and the 4th bands are formed by $p_x$ and $p_y$ orbitals centered at $A$ and $B$ points, respectively. While the 1st and 2nd bands are formed by $s_x$ and $s_y$ orbitals. With the on-site interaction, pairing occurs among orbitals at the same site but with different spin states. In the weak coupling limit, we only need to consider pairing among fermions close to the Fermi surfaces. Therefore, we mainly need to consider order parameters $\langle \hat{s}_{x,\downarrow}\hat{p}_{x,\uparrow}\rangle$ and $\langle \hat{s}_{y,\downarrow}\hat{p}_{y,\uparrow}\rangle$. Since we consider the case with well-matched spin-up and spin-down Fermi surfaces, it is reasonable to assume that the order parameter shows the same periodicity as the lattice potential. Taking the inversion center as A or B points, the order parameter shows an odd parity.
Alternatively, if we set the inversion center at the O point, the pairing order parameter still shows an odd parity due to its periodicity.

In summary, no matter what tight-binding models we used, in the weak coupling limit, well-matched spin-up and spin-down Fermi surfaces crossing the 3rd and 2nd
bands, respectively, should lead to an odd-parity pairing with inversion center defined at the O point. From a symmetry point of view, the pairing order parameter may largely inherit the point-group symmetry of the system.
Combined with the $D_4$ point-group symmetry of the lattice where the lattice center is also set at the O point, the highest symmetric pairing phase is invariant under the combined $\pm\pi/2$ gauge rotation and the $C_4$ space rotation. Our numerics confirms the existence of the anticipated highest symmetric odd-parity superfluid phases. We should note that all these results are based on the weak coupling assumption, where the order parameter should be less than other energy scales, such as band widths.

\begin{figure}[t]
\centering
\includegraphics[width=0.7\linewidth]{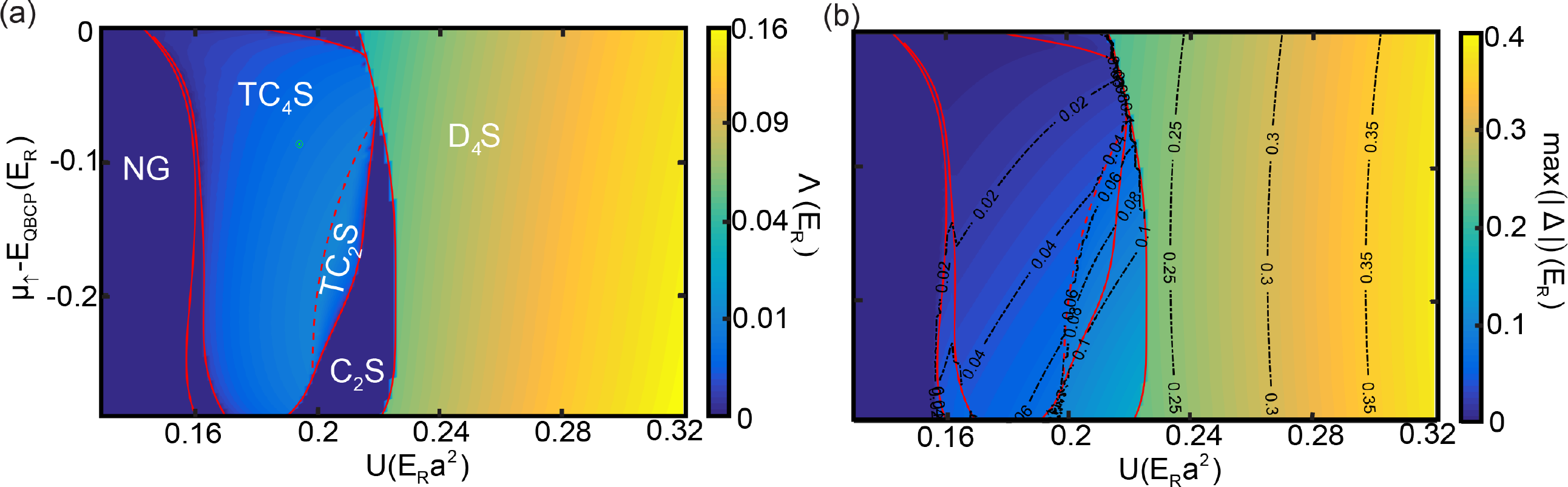}
\caption{(a) Zero-temperature ground-state phase diagram by varying the s-wave interaction and the chemical potentials with the tight-binding model II. Different from that shown in Fig.~\ref{sfig4}(b). We assume that the unit cell of the order parameter is enlarged to two unit cells of the lattice. For the special case denoted by the green $\odot$, we further perform a finite-temperature calculation. Details are shown in Fig.~\ref{sfig6}. (b) The corresponding maximum absolute value of the order parameters. Here the dotted-dash lines represent its contours.}
\label{sfig5}
\end{figure}

However, in the strong coupling limit, the assumption we used can be failed. For instance, the periodicity of the order parameter can be different from the lattice potential. In this case, we cannot infer that the highest symmetric phase should be invariant under the combined $\pm\pi/2$ gauge rotation and $C_4$ space rotation. Numerically, we relax the periodicity condition for the order parameters and assume that one unit cell of the order parameters covers two unit cells of the lattice potential. We then use the tight-binding model II and perform mean-field calculations. The ground-state phase diagram is shown in Fig.~\ref{sfig5}(a). We find that for the strong coupling limit with a stronger interaction, the ground state shows an even-parity pairing phase denoted by $\rm D_4S$ where the periodicity of the order parameter is different from the lattice. For the $\rm D_4S$ phase, the order parameters take the same $D_4$ point-group symmetry as the lattice. For instance, for the case with parameters $U=0.2279\,E_Ra^2$, $\mu_{\uparrow}=-1.4762\, E_R$, and $\mu_{\downarrow}=-4.3858\,E_R$, the ground state order parameters are
$\langle \hat{s}_{x,\mathbf{r}_{A_1},\downarrow}\hat{p}_{x,\mathbf{r}_{A_1},\uparrow}\rangle =\langle \hat{s}_{y,\mathbf{r}_{B_1},\downarrow}\hat{p}_{y,\mathbf{r}_{B_1},\uparrow}\rangle =-\langle \hat{s}_{x,\mathbf{r}_{A_2},\downarrow}\hat{p}_{x,\mathbf{r}_{A_2},\uparrow}\rangle =-\langle \hat{s}_{y,\mathbf{r}_{B_2},\downarrow}\hat{p}_{y,\mathbf{r}_{B_2},\uparrow}\rangle =-0.2398\, E_R$. Here $\mathbf{r}_{A_1}=(m+n,-m+n)a+(a/2,0)$, $\mathbf{r}_{B_1}=(m+n,-m+n)a+(0,a/2)$, $\mathbf{r}_{A_2}=(m+n,-m+n)a-(a/2,0)$, and $\mathbf{r}_{B_2}=(m+n,-m+n)a-(0,a/2)$ with $m,n\in\mathbb{Z}$.

Differently, topological phases $\rm TC_4S$ and $\rm TC_2S$ still persist. The periodicity of their order parameters is the same as the lattice. The Figure~\ref{sfig5}(b) shows the maximum absolute value of the ground-state order parameters defined in Eq.~(\ref{sop}). We can infer that for the topological superfluid phases: $\rm TC_4S$ and $\rm TC_2S$, the maximum value of the order parameters is small ($\sim (0,0.1)E_R$) comparing to other energy scales, such as the band width. This indicates that these two topological phases fall into the weak coupling regime, which validates our mean-field calculations.

\begin{figure}[H]
\centering
\includegraphics[width=0.4\linewidth]{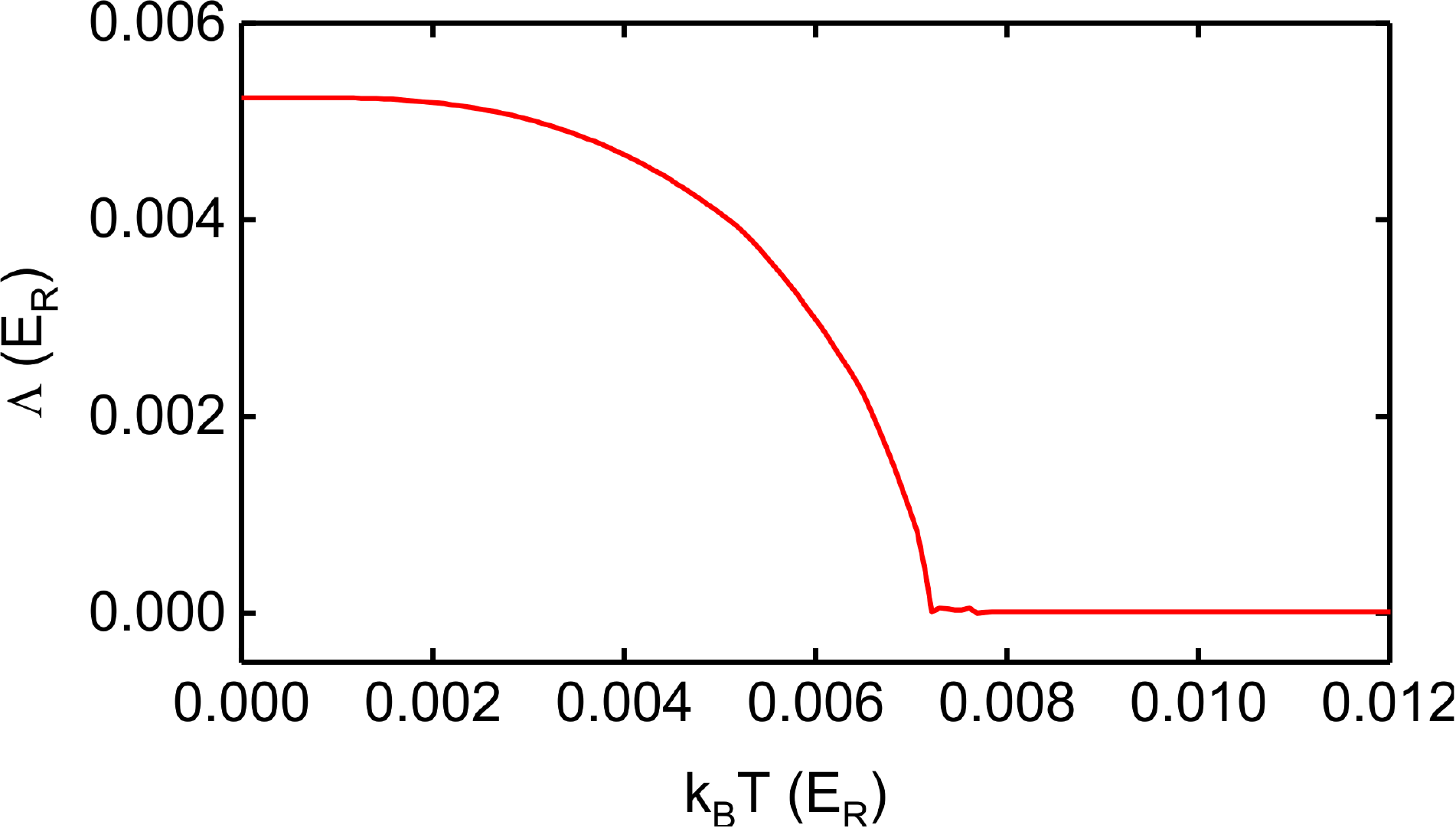}
\caption{Temperature dependent quasiparticle excitation gap $\Lambda$ for the special case with $U=0.1944\,E_Ra^2$, $\mu_{\uparrow}=-1.4762\, E_R$, and $\mu_{\downarrow}=-4.3858\,E_R$.}
\label{sfig6}
\end{figure}

Our calculations are almost performed at zero temperature. For the special case with $U=0.1944\,E_Ra^2$, $\mu_{\uparrow}=-1.4762\, E_R$, and $\mu_{\downarrow}=-4.3858\,E_R$, we also perform a finite-temperature calculation. The corresponding ground state at zero temperature falls into the topological $\rm TC_4S$ phase with a topological bulk gap $\Lambda=0.0052\,E_R$. As we increase the temperature, the bulk gap continuously decreases to zero. The predicted critical temperature is on the same scale as the bulk gap at zero temperature, such as $k_BT_c\sim \Lambda$. Base on our numerical calculation, the maximum of the full pairing gap for the topological phases is about $0.01\, E_R$, which corresponds to an experimentally feasible temperature scale of $10\,\rm nK$.


\begin{thebibliography}{43}%
\makeatletter
\providecommand \@ifxundefined [1]{%
 \@ifx{#1\undefined}
}%
\providecommand \@ifnum [1]{%
 \ifnum #1\expandafter \@firstoftwo
 \else \expandafter \@secondoftwo
 \fi
}%
\providecommand \@ifx [1]{%
 \ifx #1\expandafter \@firstoftwo
 \else \expandafter \@secondoftwo
 \fi
}%
\providecommand \natexlab [1]{#1}%
\providecommand \enquote  [1]{``#1''}%
\providecommand \bibnamefont  [1]{#1}%
\providecommand \bibfnamefont [1]{#1}%
\providecommand \citenamefont [1]{#1}%
\providecommand \href@noop [0]{\@secondoftwo}%
\providecommand \href [0]{\begingroup \@sanitize@url \@href}%
\providecommand \@href[1]{\@@startlink{#1}\@@href}%
\providecommand \@@href[1]{\endgroup#1\@@endlink}%
\providecommand \@sanitize@url [0]{\catcode `\\12\catcode `\$12\catcode
  `\&12\catcode `\#12\catcode `\^12\catcode `\_12\catcode `\%12\relax}%
\providecommand \@@startlink[1]{}%
\providecommand \@@endlink[0]{}%
\providecommand \url  [0]{\begingroup\@sanitize@url \@url }%
\providecommand \@url [1]{\endgroup\@href {#1}{\urlprefix }}%
\providecommand \urlprefix  [0]{URL }%
\providecommand \Eprint [0]{\href }%
\providecommand \doibase [0]{http://dx.doi.org/}%
\providecommand \selectlanguage [0]{\@gobble}%
\providecommand \bibinfo  [0]{\@secondoftwo}%
\providecommand \bibfield  [0]{\@secondoftwo}%
\providecommand \translation [1]{[#1]}%
\providecommand \BibitemOpen [0]{}%
\providecommand \bibitemStop [0]{}%
\providecommand \bibitemNoStop [0]{.\EOS\space}%
\providecommand \EOS [0]{\spacefactor3000\relax}%
\providecommand \BibitemShut  [1]{\csname bibitem#1\endcsname}%
\let\auto@bib@innerbib\@empty
\bibitem [{\citenamefont {Volovik}(2003)}]{Volovik2003}%
  \BibitemOpen
  \bibfield  {author} {\bibinfo {author} {\bibfnamefont {G.~E.}\ \bibnamefont
  {Volovik}},\ }\href@noop {} {\emph {\bibinfo {title} {The Universe in a
  Helium Droplet}}}\ (\bibinfo  {publisher} {Oxford University Press},\
  \bibinfo {address} {New York},\ \bibinfo {year} {2003})\BibitemShut {NoStop}%
\bibitem [{\citenamefont {Read}\ and\ \citenamefont {Green}(2000)}]{Read2000}%
  \BibitemOpen
  \bibfield  {author} {\bibinfo {author} {\bibfnamefont {N.}~\bibnamefont
  {Read}}\ and\ \bibinfo {author} {\bibfnamefont {D.}~\bibnamefont {Green}},\
  }\href {\doibase 10.1103/PhysRevB.61.10267} {\bibfield  {journal} {\bibinfo
  {journal} {Phys. Rev. B}\ }\textbf {\bibinfo {volume} {61}},\ \bibinfo
  {pages} {10267} (\bibinfo {year} {2000})}\BibitemShut {NoStop}%
\bibitem [{\citenamefont {Nayak}\ \emph {et~al.}(2008)\citenamefont {Nayak},
  \citenamefont {Simon}, \citenamefont {Stern}, \citenamefont {Freedman},\ and\
  \citenamefont {Das~Sarma}}]{Nayak2008}%
  \BibitemOpen
  \bibfield  {author} {\bibinfo {author} {\bibfnamefont {C.}~\bibnamefont
  {Nayak}}, \bibinfo {author} {\bibfnamefont {S.~H.}\ \bibnamefont {Simon}},
  \bibinfo {author} {\bibfnamefont {A.}~\bibnamefont {Stern}}, \bibinfo
  {author} {\bibfnamefont {M.}~\bibnamefont {Freedman}}, \ and\ \bibinfo
  {author} {\bibfnamefont {S.}~\bibnamefont {Das~Sarma}},\ }\href {\doibase
  10.1103/RevModPhys.80.1083} {\bibfield  {journal} {\bibinfo  {journal} {Rev.
  Mod. Phys.}\ }\textbf {\bibinfo {volume} {80}},\ \bibinfo {pages} {1083}
  (\bibinfo {year} {2008})}\BibitemShut {NoStop}%
\bibitem [{\citenamefont {Kallin}(2012)}]{Kallin2012}%
  \BibitemOpen
  \bibfield  {author} {\bibinfo {author} {\bibfnamefont {C.}~\bibnamefont
  {Kallin}},\ }\href {http://stacks.iop.org/0034-4885/75/i=4/a=042501}
  {\bibfield  {journal} {\bibinfo  {journal} {Reports on Progress in Physics}\
  }\textbf {\bibinfo {volume} {75}},\ \bibinfo {pages} {042501} (\bibinfo
  {year} {2012})}\BibitemShut {NoStop}%
\bibitem [{\citenamefont {{Sato}}\ and\ \citenamefont
  {{Ando}}(2017)}]{Sato2016}%
  \BibitemOpen
  \bibfield  {author} {\bibinfo {author} {\bibfnamefont {M.}~\bibnamefont
  {{Sato}}}\ and\ \bibinfo {author} {\bibfnamefont {Y.}~\bibnamefont
  {{Ando}}},\ }\href {\doibase 10.1088/1361-6633/aa6ac7} {\bibfield  {journal}
  {\bibinfo  {journal} {Reports on Progress in Physics}\ }\textbf {\bibinfo
  {volume} {80}},\ \bibinfo {eid} {076501} (\bibinfo {year}
  {2017})}\BibitemShut {NoStop}%
\bibitem [{\citenamefont {{Gurarie}}\ and\ \citenamefont
  {{Radzihovsky}}(2007)}]{Gurarie2007}%
  \BibitemOpen
  \bibfield  {author} {\bibinfo {author} {\bibfnamefont {V.}~\bibnamefont
  {{Gurarie}}}\ and\ \bibinfo {author} {\bibfnamefont {L.}~\bibnamefont
  {{Radzihovsky}}},\ }\href {\doibase 10.1016/j.aop.2006.10.009} {\bibfield
  {journal} {\bibinfo  {journal} {Annals of Physics}\ }\textbf {\bibinfo
  {volume} {322}},\ \bibinfo {pages} {2} (\bibinfo {year} {2007})}\BibitemShut
  {NoStop}%
\bibitem [{\citenamefont {Zhang}\ \emph {et~al.}(2008)\citenamefont {Zhang},
  \citenamefont {Tewari}, \citenamefont {Lutchyn},\ and\ \citenamefont
  {Das~Sarma}}]{Zhang2008}%
  \BibitemOpen
  \bibfield  {author} {\bibinfo {author} {\bibfnamefont {C.}~\bibnamefont
  {Zhang}}, \bibinfo {author} {\bibfnamefont {S.}~\bibnamefont {Tewari}},
  \bibinfo {author} {\bibfnamefont {R.~M.}\ \bibnamefont {Lutchyn}}, \ and\
  \bibinfo {author} {\bibfnamefont {S.}~\bibnamefont {Das~Sarma}},\ }\href
  {\doibase 10.1103/PhysRevLett.101.160401} {\bibfield  {journal} {\bibinfo
  {journal} {Phys. Rev. Lett.}\ }\textbf {\bibinfo {volume} {101}},\ \bibinfo
  {pages} {160401} (\bibinfo {year} {2008})}\BibitemShut {NoStop}%
\bibitem [{\citenamefont {Sato}\ \emph {et~al.}(2009)\citenamefont {Sato},
  \citenamefont {Takahashi},\ and\ \citenamefont {Fujimoto}}]{Sato2009}%
  \BibitemOpen
  \bibfield  {author} {\bibinfo {author} {\bibfnamefont {M.}~\bibnamefont
  {Sato}}, \bibinfo {author} {\bibfnamefont {Y.}~\bibnamefont {Takahashi}}, \
  and\ \bibinfo {author} {\bibfnamefont {S.}~\bibnamefont {Fujimoto}},\ }\href
  {\doibase 10.1103/PhysRevLett.103.020401} {\bibfield  {journal} {\bibinfo
  {journal} {Phys. Rev. Lett.}\ }\textbf {\bibinfo {volume} {103}},\ \bibinfo
  {pages} {020401} (\bibinfo {year} {2009})}\BibitemShut {NoStop}%
\bibitem [{\citenamefont {Cooper}\ and\ \citenamefont
  {Shlyapnikov}(2009)}]{Cooper2009}%
  \BibitemOpen
  \bibfield  {author} {\bibinfo {author} {\bibfnamefont {N.~R.}\ \bibnamefont
  {Cooper}}\ and\ \bibinfo {author} {\bibfnamefont {G.~V.}\ \bibnamefont
  {Shlyapnikov}},\ }\href {\doibase 10.1103/PhysRevLett.103.155302} {\bibfield
  {journal} {\bibinfo  {journal} {Phys. Rev. Lett.}\ }\textbf {\bibinfo
  {volume} {103}},\ \bibinfo {pages} {155302} (\bibinfo {year}
  {2009})}\BibitemShut {NoStop}%
\bibitem [{\citenamefont {Liu}\ \emph {et~al.}(2014{\natexlab{a}})\citenamefont
  {Liu}, \citenamefont {Law},\ and\ \citenamefont {Ng}}]{Liu2014XJL}%
  \BibitemOpen
  \bibfield  {author} {\bibinfo {author} {\bibfnamefont {X.-J.}\ \bibnamefont
  {Liu}}, \bibinfo {author} {\bibfnamefont {K.~T.}\ \bibnamefont {Law}}, \ and\
  \bibinfo {author} {\bibfnamefont {T.~K.}\ \bibnamefont {Ng}},\ }\href
  {\doibase 10.1103/PhysRevLett.112.086401} {\bibfield  {journal} {\bibinfo
  {journal} {Phys. Rev. Lett.}\ }\textbf {\bibinfo {volume} {112}},\ \bibinfo
  {pages} {086401} (\bibinfo {year} {2014}{\natexlab{a}})}\BibitemShut
  {NoStop}%
\bibitem [{\citenamefont {{B{\"u}hler}}\ \emph {et~al.}(2014)\citenamefont
  {{B{\"u}hler}}, \citenamefont {{Lang}}, \citenamefont {{Kraus}},
  \citenamefont {{M{\"o}ller}}, \citenamefont {{Huber}},\ and\ \citenamefont
  {{B{\"u}chler}}}]{Buhler2014}%
  \BibitemOpen
  \bibfield  {author} {\bibinfo {author} {\bibfnamefont {A.}~\bibnamefont
  {{B{\"u}hler}}}, \bibinfo {author} {\bibfnamefont {N.}~\bibnamefont
  {{Lang}}}, \bibinfo {author} {\bibfnamefont {C.~V.}\ \bibnamefont {{Kraus}}},
  \bibinfo {author} {\bibfnamefont {G.}~\bibnamefont {{M{\"o}ller}}}, \bibinfo
  {author} {\bibfnamefont {S.~D.}\ \bibnamefont {{Huber}}}, \ and\ \bibinfo
  {author} {\bibfnamefont {H.~P.}\ \bibnamefont {{B{\"u}chler}}},\ }\href
  {\doibase 10.1038/ncomms5504} {\bibfield  {journal} {\bibinfo  {journal}
  {Nature Communications}\ }\textbf {\bibinfo {volume} {5}},\ \bibinfo {eid}
  {4504} (\bibinfo {year} {2014})}\BibitemShut {NoStop}%
\bibitem [{\citenamefont {Liu}\ \emph {et~al.}(2014{\natexlab{b}})\citenamefont
  {Liu}, \citenamefont {Li}, \citenamefont {Wu},\ and\ \citenamefont
  {Liu}}]{Liu2014}%
  \BibitemOpen
  \bibfield  {author} {\bibinfo {author} {\bibfnamefont {B.}~\bibnamefont
  {Liu}}, \bibinfo {author} {\bibfnamefont {X.}~\bibnamefont {Li}}, \bibinfo
  {author} {\bibfnamefont {B.}~\bibnamefont {Wu}}, \ and\ \bibinfo {author}
  {\bibfnamefont {W.~V.}\ \bibnamefont {Liu}},\ }\href {\doibase
  10.1038/ncomms6064} {\bibfield  {journal} {\bibinfo  {journal} {Nature
  Communications}\ }\textbf {\bibinfo {volume} {5}},\ \bibinfo {pages} {5064}
  (\bibinfo {year} {2014}{\natexlab{b}})}\BibitemShut {NoStop}%
\bibitem [{\citenamefont {Zhang}\ \emph {et~al.}(2015)\citenamefont {Zhang},
  \citenamefont {Lang},\ and\ \citenamefont {Zhou}}]{Zhang2015}%
  \BibitemOpen
  \bibfield  {author} {\bibinfo {author} {\bibfnamefont {S.-L.}\ \bibnamefont
  {Zhang}}, \bibinfo {author} {\bibfnamefont {L.-J.}\ \bibnamefont {Lang}}, \
  and\ \bibinfo {author} {\bibfnamefont {Q.}~\bibnamefont {Zhou}},\ }\href
  {\doibase 10.1103/PhysRevLett.115.225301} {\bibfield  {journal} {\bibinfo
  {journal} {Phys. Rev. Lett.}\ }\textbf {\bibinfo {volume} {115}},\ \bibinfo
  {pages} {225301} (\bibinfo {year} {2015})}\BibitemShut {NoStop}%
\bibitem [{\citenamefont {Wang}\ \emph {et~al.}(2016)\citenamefont {Wang},
  \citenamefont {Zheng}, \citenamefont {Pu}, \citenamefont {Zou},\ and\
  \citenamefont {Guo}}]{Wang2016}%
  \BibitemOpen
  \bibfield  {author} {\bibinfo {author} {\bibfnamefont {B.}~\bibnamefont
  {Wang}}, \bibinfo {author} {\bibfnamefont {Z.}~\bibnamefont {Zheng}},
  \bibinfo {author} {\bibfnamefont {H.}~\bibnamefont {Pu}}, \bibinfo {author}
  {\bibfnamefont {X.}~\bibnamefont {Zou}}, \ and\ \bibinfo {author}
  {\bibfnamefont {G.}~\bibnamefont {Guo}},\ }\href {\doibase
  10.1103/PhysRevA.93.031602} {\bibfield  {journal} {\bibinfo  {journal} {Phys.
  Rev. A}\ }\textbf {\bibinfo {volume} {93}},\ \bibinfo {pages} {031602}
  (\bibinfo {year} {2016})}\BibitemShut {NoStop}%
\bibitem [{\citenamefont {Wu}\ and\ \citenamefont {Bruun}(2016)}]{Wu2016}%
  \BibitemOpen
  \bibfield  {author} {\bibinfo {author} {\bibfnamefont {Z.}~\bibnamefont
  {Wu}}\ and\ \bibinfo {author} {\bibfnamefont {G.~M.}\ \bibnamefont {Bruun}},\
  }\href {\doibase 10.1103/PhysRevLett.117.245302} {\bibfield  {journal}
  {\bibinfo  {journal} {Phys. Rev. Lett.}\ }\textbf {\bibinfo {volume} {117}},\
  \bibinfo {pages} {245302} (\bibinfo {year} {2016})}\BibitemShut {NoStop}%
\bibitem [{\citenamefont {Fu}\ and\ \citenamefont {Kane}(2008)}]{Fu2008}%
  \BibitemOpen
  \bibfield  {author} {\bibinfo {author} {\bibfnamefont {L.}~\bibnamefont
  {Fu}}\ and\ \bibinfo {author} {\bibfnamefont {C.~L.}\ \bibnamefont {Kane}},\
  }\href {\doibase 10.1103/PhysRevLett.100.096407} {\bibfield  {journal}
  {\bibinfo  {journal} {Phys. Rev. Lett.}\ }\textbf {\bibinfo {volume} {100}},\
  \bibinfo {pages} {096407} (\bibinfo {year} {2008})}\BibitemShut {NoStop}%
\bibitem [{\citenamefont {{Wang}}\ \emph {et~al.}(2012)\citenamefont {{Wang}},
  \citenamefont {{Liu}}, \citenamefont {{Xu}}, \citenamefont {{Yang}},
  \citenamefont {{Miao}}, \citenamefont {{Yao}}, \citenamefont {{Gao}},
  \citenamefont {{Shen}}, \citenamefont {{Ma}}, \citenamefont {{Chen}},
  \citenamefont {{Xu}}, \citenamefont {{Liu}}, \citenamefont {{Zhang}},
  \citenamefont {{Qian}}, \citenamefont {{Jia}},\ and\ \citenamefont
  {{Xue}}}]{Jia2012}%
  \BibitemOpen
  \bibfield  {author} {\bibinfo {author} {\bibfnamefont {M.-X.}\ \bibnamefont
  {{Wang}}}, \bibinfo {author} {\bibfnamefont {C.}~\bibnamefont {{Liu}}},
  \bibinfo {author} {\bibfnamefont {J.-P.}\ \bibnamefont {{Xu}}}, \bibinfo
  {author} {\bibfnamefont {F.}~\bibnamefont {{Yang}}}, \bibinfo {author}
  {\bibfnamefont {L.}~\bibnamefont {{Miao}}}, \bibinfo {author} {\bibfnamefont
  {M.-Y.}\ \bibnamefont {{Yao}}}, \bibinfo {author} {\bibfnamefont {C.~L.}\
  \bibnamefont {{Gao}}}, \bibinfo {author} {\bibfnamefont {C.}~\bibnamefont
  {{Shen}}}, \bibinfo {author} {\bibfnamefont {X.}~\bibnamefont {{Ma}}},
  \bibinfo {author} {\bibfnamefont {X.}~\bibnamefont {{Chen}}}, \bibinfo
  {author} {\bibfnamefont {Z.-A.}\ \bibnamefont {{Xu}}}, \bibinfo {author}
  {\bibfnamefont {Y.}~\bibnamefont {{Liu}}}, \bibinfo {author} {\bibfnamefont
  {S.-C.}\ \bibnamefont {{Zhang}}}, \bibinfo {author} {\bibfnamefont
  {D.}~\bibnamefont {{Qian}}}, \bibinfo {author} {\bibfnamefont {J.-F.}\
  \bibnamefont {{Jia}}}, \ and\ \bibinfo {author} {\bibfnamefont {Q.-K.}\
  \bibnamefont {{Xue}}},\ }\href {\doibase 10.1126/science.1216466} {\bibfield
  {journal} {\bibinfo  {journal} {Science}\ }\textbf {\bibinfo {volume}
  {336}},\ \bibinfo {pages} {52} (\bibinfo {year} {2012})}\BibitemShut
  {NoStop}%
\bibitem [{\citenamefont {Xu}\ \emph {et~al.}(2015)\citenamefont {Xu},
  \citenamefont {Wang}, \citenamefont {Liu}, \citenamefont {Ge}, \citenamefont
  {Yang}, \citenamefont {Liu}, \citenamefont {Xu}, \citenamefont {Guan},
  \citenamefont {Gao}, \citenamefont {Qian}, \citenamefont {Liu}, \citenamefont
  {Wang}, \citenamefont {Zhang}, \citenamefont {Xue},\ and\ \citenamefont
  {Jia}}]{Jia2015}%
  \BibitemOpen
  \bibfield  {author} {\bibinfo {author} {\bibfnamefont {J.-P.}\ \bibnamefont
  {Xu}}, \bibinfo {author} {\bibfnamefont {M.-X.}\ \bibnamefont {Wang}},
  \bibinfo {author} {\bibfnamefont {Z.~L.}\ \bibnamefont {Liu}}, \bibinfo
  {author} {\bibfnamefont {J.-F.}\ \bibnamefont {Ge}}, \bibinfo {author}
  {\bibfnamefont {X.}~\bibnamefont {Yang}}, \bibinfo {author} {\bibfnamefont
  {C.}~\bibnamefont {Liu}}, \bibinfo {author} {\bibfnamefont {Z.~A.}\
  \bibnamefont {Xu}}, \bibinfo {author} {\bibfnamefont {D.}~\bibnamefont
  {Guan}}, \bibinfo {author} {\bibfnamefont {C.~L.}\ \bibnamefont {Gao}},
  \bibinfo {author} {\bibfnamefont {D.}~\bibnamefont {Qian}}, \bibinfo {author}
  {\bibfnamefont {Y.}~\bibnamefont {Liu}}, \bibinfo {author} {\bibfnamefont
  {Q.-H.}\ \bibnamefont {Wang}}, \bibinfo {author} {\bibfnamefont {F.-C.}\
  \bibnamefont {Zhang}}, \bibinfo {author} {\bibfnamefont {Q.-K.}\ \bibnamefont
  {Xue}}, \ and\ \bibinfo {author} {\bibfnamefont {J.-F.}\ \bibnamefont
  {Jia}},\ }\href {\doibase 10.1103/PhysRevLett.114.017001} {\bibfield
  {journal} {\bibinfo  {journal} {Phys. Rev. Lett.}\ }\textbf {\bibinfo
  {volume} {114}},\ \bibinfo {pages} {017001} (\bibinfo {year}
  {2015})}\BibitemShut {NoStop}%
\bibitem [{\citenamefont {Maeno}\ \emph {et~al.}(1994)\citenamefont {Maeno},
  \citenamefont {Hashimoto}, \citenamefont {Yoshida}, \citenamefont
  {Nishizaki}, \citenamefont {Fujita}, \citenamefont {Bednorz},\ and\
  \citenamefont {Lichtenberg}}]{Maeno1994}%
  \BibitemOpen
  \bibfield  {author} {\bibinfo {author} {\bibfnamefont {Y.}~\bibnamefont
  {Maeno}}, \bibinfo {author} {\bibfnamefont {H.}~\bibnamefont {Hashimoto}},
  \bibinfo {author} {\bibfnamefont {K.}~\bibnamefont {Yoshida}}, \bibinfo
  {author} {\bibfnamefont {S.}~\bibnamefont {Nishizaki}}, \bibinfo {author}
  {\bibfnamefont {T.}~\bibnamefont {Fujita}}, \bibinfo {author} {\bibfnamefont
  {J.~G.}\ \bibnamefont {Bednorz}}, \ and\ \bibinfo {author} {\bibfnamefont
  {F.}~\bibnamefont {Lichtenberg}},\ }\href
  {http://dx.doi.org/10.1038/372532a0} {\bibfield  {journal} {\bibinfo
  {journal} {\nat}\ }\textbf {\bibinfo {volume} {372}},\ \bibinfo {pages} {532}
  (\bibinfo {year} {1994})}\BibitemShut {NoStop}%
\bibitem [{\citenamefont {Mackenzie}\ and\ \citenamefont
  {Maeno}(2003)}]{Mackenzie2003}%
  \BibitemOpen
  \bibfield  {author} {\bibinfo {author} {\bibfnamefont {A.~P.}\ \bibnamefont
  {Mackenzie}}\ and\ \bibinfo {author} {\bibfnamefont {Y.}~\bibnamefont
  {Maeno}},\ }\href {\doibase 10.1103/RevModPhys.75.657} {\bibfield  {journal}
  {\bibinfo  {journal} {Rev. Mod. Phys.}\ }\textbf {\bibinfo {volume} {75}},\
  \bibinfo {pages} {657} (\bibinfo {year} {2003})}\BibitemShut {NoStop}%
\bibitem [{\citenamefont {{Soltan-Panahi}}\ \emph {et~al.}(2011)\citenamefont
  {{Soltan-Panahi}}, \citenamefont {{Struck}}, \citenamefont {{Hauke}},
  \citenamefont {{Bick}}, \citenamefont {{Plenkers}}, \citenamefont
  {{Meineke}}, \citenamefont {{Becker}}, \citenamefont {{Windpassinger}},
  \citenamefont {{Lewenstein}},\ and\ \citenamefont
  {{Sengstock}}}]{Sengstock2010}%
  \BibitemOpen
  \bibfield  {author} {\bibinfo {author} {\bibfnamefont {P.}~\bibnamefont
  {{Soltan-Panahi}}}, \bibinfo {author} {\bibfnamefont {J.}~\bibnamefont
  {{Struck}}}, \bibinfo {author} {\bibfnamefont {P.}~\bibnamefont {{Hauke}}},
  \bibinfo {author} {\bibfnamefont {A.}~\bibnamefont {{Bick}}}, \bibinfo
  {author} {\bibfnamefont {W.}~\bibnamefont {{Plenkers}}}, \bibinfo {author}
  {\bibfnamefont {G.}~\bibnamefont {{Meineke}}}, \bibinfo {author}
  {\bibfnamefont {C.}~\bibnamefont {{Becker}}}, \bibinfo {author}
  {\bibfnamefont {P.}~\bibnamefont {{Windpassinger}}}, \bibinfo {author}
  {\bibfnamefont {M.}~\bibnamefont {{Lewenstein}}}, \ and\ \bibinfo {author}
  {\bibfnamefont {K.}~\bibnamefont {{Sengstock}}},\ }\href {\doibase
  10.1038/nphys1916} {\bibfield  {journal} {\bibinfo  {journal} {Nature
  Physics}\ }\textbf {\bibinfo {volume} {7}},\ \bibinfo {pages} {434} (\bibinfo
  {year} {2011})}\BibitemShut {NoStop}%
\bibitem [{\citenamefont {Jo}\ \emph {et~al.}(2012)\citenamefont {Jo},
  \citenamefont {Guzman}, \citenamefont {Thomas}, \citenamefont {Hosur},
  \citenamefont {Vishwanath},\ and\ \citenamefont {Stamper-Kurn}}]{Jo2012}%
  \BibitemOpen
  \bibfield  {author} {\bibinfo {author} {\bibfnamefont {G.-B.}\ \bibnamefont
  {Jo}}, \bibinfo {author} {\bibfnamefont {J.}~\bibnamefont {Guzman}}, \bibinfo
  {author} {\bibfnamefont {C.~K.}\ \bibnamefont {Thomas}}, \bibinfo {author}
  {\bibfnamefont {P.}~\bibnamefont {Hosur}}, \bibinfo {author} {\bibfnamefont
  {A.}~\bibnamefont {Vishwanath}}, \ and\ \bibinfo {author} {\bibfnamefont
  {D.~M.}\ \bibnamefont {Stamper-Kurn}},\ }\href {\doibase
  10.1103/PhysRevLett.108.045305} {\bibfield  {journal} {\bibinfo  {journal}
  {Phys. Rev. Lett.}\ }\textbf {\bibinfo {volume} {108}},\ \bibinfo {pages}
  {045305} (\bibinfo {year} {2012})}\BibitemShut {NoStop}%
\bibitem [{\citenamefont {Taie}\ \emph {et~al.}(2015)\citenamefont {Taie},
  \citenamefont {Ozawa}, \citenamefont {Ichinose}, \citenamefont {Nishio},
  \citenamefont {Nakajima},\ and\ \citenamefont {Takahashi}}]{Taiee2015}%
  \BibitemOpen
  \bibfield  {author} {\bibinfo {author} {\bibfnamefont {S.}~\bibnamefont
  {Taie}}, \bibinfo {author} {\bibfnamefont {H.}~\bibnamefont {Ozawa}},
  \bibinfo {author} {\bibfnamefont {T.}~\bibnamefont {Ichinose}}, \bibinfo
  {author} {\bibfnamefont {T.}~\bibnamefont {Nishio}}, \bibinfo {author}
  {\bibfnamefont {S.}~\bibnamefont {Nakajima}}, \ and\ \bibinfo {author}
  {\bibfnamefont {Y.}~\bibnamefont {Takahashi}},\ }\href {\doibase
  10.1126/sciadv.1500854} {\bibfield  {journal} {\bibinfo  {journal} {Science
  Advances}\ }\textbf {\bibinfo {volume} {1}},\ \bibinfo {pages} {e1500854}
  (\bibinfo {year} {2015})}\BibitemShut {NoStop}%
\bibitem [{\citenamefont {Sebby-Strabley}\ \emph {et~al.}(2006)\citenamefont
  {Sebby-Strabley}, \citenamefont {Anderlini}, \citenamefont {Jessen},\ and\
  \citenamefont {Porto}}]{Sebby-Strabley2006}%
  \BibitemOpen
  \bibfield  {author} {\bibinfo {author} {\bibfnamefont {J.}~\bibnamefont
  {Sebby-Strabley}}, \bibinfo {author} {\bibfnamefont {M.}~\bibnamefont
  {Anderlini}}, \bibinfo {author} {\bibfnamefont {P.~S.}\ \bibnamefont
  {Jessen}}, \ and\ \bibinfo {author} {\bibfnamefont {J.~V.}\ \bibnamefont
  {Porto}},\ }\href {\doibase 10.1103/PhysRevA.73.033605} {\bibfield  {journal}
  {\bibinfo  {journal} {Phys. Rev. A}\ }\textbf {\bibinfo {volume} {73}},\
  \bibinfo {pages} {033605} (\bibinfo {year} {2006})}\BibitemShut {NoStop}%
\bibitem [{\citenamefont {{Wirth}}\ \emph {et~al.}(2011)\citenamefont
  {{Wirth}}, \citenamefont {{{\"O}lschl{\"a}ger}},\ and\ \citenamefont
  {{Hemmerich}}}]{Wirth2011}%
  \BibitemOpen
  \bibfield  {author} {\bibinfo {author} {\bibfnamefont {G.}~\bibnamefont
  {{Wirth}}}, \bibinfo {author} {\bibfnamefont {M.}~\bibnamefont
  {{{\"O}lschl{\"a}ger}}}, \ and\ \bibinfo {author} {\bibfnamefont
  {A.}~\bibnamefont {{Hemmerich}}},\ }\href {\doibase 10.1038/nphys1857}
  {\bibfield  {journal} {\bibinfo  {journal} {Nature Physics}\ }\textbf
  {\bibinfo {volume} {7}},\ \bibinfo {pages} {147} (\bibinfo {year}
  {2011})}\BibitemShut {NoStop}%
\bibitem [{\citenamefont {{Sun}}\ \emph {et~al.}(2012)\citenamefont {{Sun}},
  \citenamefont {{Liu}}, \citenamefont {{Hemmerich}},\ and\ \citenamefont {{Das
  Sarma}}}]{Sun2012}%
  \BibitemOpen
  \bibfield  {author} {\bibinfo {author} {\bibfnamefont {K.}~\bibnamefont
  {{Sun}}}, \bibinfo {author} {\bibfnamefont {W.~V.}\ \bibnamefont {{Liu}}},
  \bibinfo {author} {\bibfnamefont {A.}~\bibnamefont {{Hemmerich}}}, \ and\
  \bibinfo {author} {\bibfnamefont {S.}~\bibnamefont {{Das Sarma}}},\ }\href
  {\doibase 10.1038/nphys2134} {\bibfield  {journal} {\bibinfo  {journal}
  {Nature Physics}\ }\textbf {\bibinfo {volume} {8}},\ \bibinfo {pages} {67}
  (\bibinfo {year} {2012})}\BibitemShut {NoStop}%
\bibitem [{\citenamefont {Sun}\ \emph {et~al.}(2009)\citenamefont {Sun},
  \citenamefont {Yao}, \citenamefont {Fradkin},\ and\ \citenamefont
  {Kivelson}}]{Sun2009}%
  \BibitemOpen
  \bibfield  {author} {\bibinfo {author} {\bibfnamefont {K.}~\bibnamefont
  {Sun}}, \bibinfo {author} {\bibfnamefont {H.}~\bibnamefont {Yao}}, \bibinfo
  {author} {\bibfnamefont {E.}~\bibnamefont {Fradkin}}, \ and\ \bibinfo
  {author} {\bibfnamefont {S.~A.}\ \bibnamefont {Kivelson}},\ }\href {\doibase
  10.1103/PhysRevLett.103.046811} {\bibfield  {journal} {\bibinfo  {journal}
  {Phys. Rev. Lett.}\ }\textbf {\bibinfo {volume} {103}},\ \bibinfo {pages}
  {046811} (\bibinfo {year} {2009})}\BibitemShut {NoStop}%
\bibitem [{\citenamefont {Zwierlein}\ \emph {et~al.}(2006)\citenamefont
  {Zwierlein}, \citenamefont {Schirotzek}, \citenamefont {Schunck},\ and\
  \citenamefont {Ketterle}}]{Zwierlein2006}%
  \BibitemOpen
  \bibfield  {author} {\bibinfo {author} {\bibfnamefont {M.~W.}\ \bibnamefont
  {Zwierlein}}, \bibinfo {author} {\bibfnamefont {A.}~\bibnamefont
  {Schirotzek}}, \bibinfo {author} {\bibfnamefont {C.~H.}\ \bibnamefont
  {Schunck}}, \ and\ \bibinfo {author} {\bibfnamefont {W.}~\bibnamefont
  {Ketterle}},\ }\href {\doibase 10.1126/science.1122318} {\bibfield  {journal}
  {\bibinfo  {journal} {Science}\ }\textbf {\bibinfo {volume} {311}},\ \bibinfo
  {pages} {492} (\bibinfo {year} {2006})}\BibitemShut {NoStop}%
\bibitem [{\citenamefont {Partridge}\ \emph {et~al.}(2006)\citenamefont
  {Partridge}, \citenamefont {Li}, \citenamefont {Kamar}, \citenamefont
  {Liao},\ and\ \citenamefont {Hulet}}]{Partridge2006}%
  \BibitemOpen
  \bibfield  {author} {\bibinfo {author} {\bibfnamefont {G.~B.}\ \bibnamefont
  {Partridge}}, \bibinfo {author} {\bibfnamefont {W.}~\bibnamefont {Li}},
  \bibinfo {author} {\bibfnamefont {R.~I.}\ \bibnamefont {Kamar}}, \bibinfo
  {author} {\bibfnamefont {Y.-a.}\ \bibnamefont {Liao}}, \ and\ \bibinfo
  {author} {\bibfnamefont {R.~G.}\ \bibnamefont {Hulet}},\ }\href {\doibase
  10.1126/science.1122876} {\bibfield  {journal} {\bibinfo  {journal}
  {Science}\ }\textbf {\bibinfo {volume} {311}},\ \bibinfo {pages} {503}
  (\bibinfo {year} {2006})}\BibitemShut {NoStop}%
\bibitem [{\citenamefont {{Chin}}\ \emph {et~al.}(2006)\citenamefont {{Chin}},
  \citenamefont {{Miller}}, \citenamefont {{Liu}}, \citenamefont {{Stan}},
  \citenamefont {{Setiawan}}, \citenamefont {{Sanner}}, \citenamefont {{Xu}},\
  and\ \citenamefont {{Ketterle}}}]{Chin2006}%
  \BibitemOpen
  \bibfield  {author} {\bibinfo {author} {\bibfnamefont {J.~K.}\ \bibnamefont
  {{Chin}}}, \bibinfo {author} {\bibfnamefont {D.~E.}\ \bibnamefont
  {{Miller}}}, \bibinfo {author} {\bibfnamefont {Y.}~\bibnamefont {{Liu}}},
  \bibinfo {author} {\bibfnamefont {C.}~\bibnamefont {{Stan}}}, \bibinfo
  {author} {\bibfnamefont {W.}~\bibnamefont {{Setiawan}}}, \bibinfo {author}
  {\bibfnamefont {C.}~\bibnamefont {{Sanner}}}, \bibinfo {author}
  {\bibfnamefont {K.}~\bibnamefont {{Xu}}}, \ and\ \bibinfo {author}
  {\bibfnamefont {W.}~\bibnamefont {{Ketterle}}},\ }\href {\doibase
  10.1038/nature05224} {\bibfield  {journal} {\bibinfo  {journal} {\nat}\
  }\textbf {\bibinfo {volume} {443}},\ \bibinfo {pages} {961} (\bibinfo {year}
  {2006})}\BibitemShut {NoStop}%
\bibitem [{\citenamefont {Zhai}\ and\ \citenamefont {Ho}(2007)}]{Zhai2007}%
  \BibitemOpen
  \bibfield  {author} {\bibinfo {author} {\bibfnamefont {H.}~\bibnamefont
  {Zhai}}\ and\ \bibinfo {author} {\bibfnamefont {T.-L.}\ \bibnamefont {Ho}},\
  }\href {\doibase 10.1103/PhysRevLett.99.100402} {\bibfield  {journal}
  {\bibinfo  {journal} {Phys. Rev. Lett.}\ }\textbf {\bibinfo {volume} {99}},\
  \bibinfo {pages} {100402} (\bibinfo {year} {2007})}\BibitemShut {NoStop}%
\bibitem [{\citenamefont {Moon}\ \emph {et~al.}(2007)\citenamefont {Moon},
  \citenamefont {Nikoli\ifmmode~\acute{c}\else \'{c}\fi{}},\ and\ \citenamefont
  {Sachdev}}]{Moon2007}%
  \BibitemOpen
  \bibfield  {author} {\bibinfo {author} {\bibfnamefont {E.~G.}\ \bibnamefont
  {Moon}}, \bibinfo {author} {\bibfnamefont {P.}~\bibnamefont
  {Nikoli\ifmmode~\acute{c}\else \'{c}\fi{}}}, \ and\ \bibinfo {author}
  {\bibfnamefont {S.}~\bibnamefont {Sachdev}},\ }\href {\doibase
  10.1103/PhysRevLett.99.230403} {\bibfield  {journal} {\bibinfo  {journal}
  {Phys. Rev. Lett.}\ }\textbf {\bibinfo {volume} {99}},\ \bibinfo {pages}
  {230403} (\bibinfo {year} {2007})}\BibitemShut {NoStop}%
\bibitem [{\citenamefont {Fulde}\ and\ \citenamefont
  {Ferrell}(1964)}]{Fulde1964}%
  \BibitemOpen
  \bibfield  {author} {\bibinfo {author} {\bibfnamefont {P.}~\bibnamefont
  {Fulde}}\ and\ \bibinfo {author} {\bibfnamefont {R.~A.}\ \bibnamefont
  {Ferrell}},\ }\href {\doibase 10.1103/PhysRev.135.A550} {\bibfield  {journal}
  {\bibinfo  {journal} {Phys. Rev.}\ }\textbf {\bibinfo {volume} {135}},\
  \bibinfo {pages} {A550} (\bibinfo {year} {1964})}\BibitemShut {NoStop}%
\bibitem [{\citenamefont {Larkin}\ and\ \citenamefont
  {Ovchinnikov}(1964)}]{Larkin1964}%
  \BibitemOpen
  \bibfield  {author} {\bibinfo {author} {\bibfnamefont {A.~I.}\ \bibnamefont
  {Larkin}}\ and\ \bibinfo {author} {\bibfnamefont {Y.~N.}\ \bibnamefont
  {Ovchinnikov}},\ }\href@noop {} {\bibfield  {journal} {\bibinfo  {journal}
  {Zh. Eksp. Teor. Fiz.}\ }\textbf {\bibinfo {volume} {47}},\ \bibinfo {pages}
  {1136} (\bibinfo {year} {1964})}\BibitemShut {NoStop}%
\bibitem [{\citenamefont {{Liao}}\ \emph {et~al.}(2010)\citenamefont {{Liao}},
  \citenamefont {{Rittner}}, \citenamefont {{Paprotta}}, \citenamefont {{Li}},
  \citenamefont {{Partridge}}, \citenamefont {{Hulet}}, \citenamefont
  {{Baur}},\ and\ \citenamefont {{Mueller}}}]{Liao2010}%
  \BibitemOpen
  \bibfield  {author} {\bibinfo {author} {\bibfnamefont {Y.-A.}\ \bibnamefont
  {{Liao}}}, \bibinfo {author} {\bibfnamefont {A.~S.~C.}\ \bibnamefont
  {{Rittner}}}, \bibinfo {author} {\bibfnamefont {T.}~\bibnamefont
  {{Paprotta}}}, \bibinfo {author} {\bibfnamefont {W.}~\bibnamefont {{Li}}},
  \bibinfo {author} {\bibfnamefont {G.~B.}\ \bibnamefont {{Partridge}}},
  \bibinfo {author} {\bibfnamefont {R.~G.}\ \bibnamefont {{Hulet}}}, \bibinfo
  {author} {\bibfnamefont {S.~K.}\ \bibnamefont {{Baur}}}, \ and\ \bibinfo
  {author} {\bibfnamefont {E.~J.}\ \bibnamefont {{Mueller}}},\ }\href {\doibase
  10.1038/nature09393} {\bibfield  {journal} {\bibinfo  {journal} {\nat}\
  }\textbf {\bibinfo {volume} {467}},\ \bibinfo {pages} {567} (\bibinfo {year}
  {2010})}\BibitemShut {NoStop}%
\bibitem [{sup()}]{suppl}%
  \BibitemOpen
  \href@noop {} {}\bibinfo {note} {Supplemental material, which includes
  Refs.~\cite{Kivelson1982,Uehlinger2013}, for additional details on
  diagonalization of the mean-field Hamiltonian, tight-binding models and
  superfluid phases calculated from the tight-binding models.}\BibitemShut
  {Stop}%
\bibitem [{\citenamefont {Xu}\ \emph {et~al.}(2016)\citenamefont {Xu},
  \citenamefont {You}, \citenamefont {Hemmerich},\ and\ \citenamefont
  {Liu}}]{Xu2016}%
  \BibitemOpen
  \bibfield  {author} {\bibinfo {author} {\bibfnamefont {Z.-F.}\ \bibnamefont
  {Xu}}, \bibinfo {author} {\bibfnamefont {L.}~\bibnamefont {You}}, \bibinfo
  {author} {\bibfnamefont {A.}~\bibnamefont {Hemmerich}}, \ and\ \bibinfo
  {author} {\bibfnamefont {W.~V.}\ \bibnamefont {Liu}},\ }\href {\doibase
  10.1103/PhysRevLett.117.085301} {\bibfield  {journal} {\bibinfo  {journal}
  {Phys. Rev. Lett.}\ }\textbf {\bibinfo {volume} {117}},\ \bibinfo {pages}
  {085301} (\bibinfo {year} {2016})}\BibitemShut {NoStop}%
\bibitem [{\citenamefont {Fu}\ and\ \citenamefont {Berg}(2010)}]{Fu2010}%
  \BibitemOpen
  \bibfield  {author} {\bibinfo {author} {\bibfnamefont {L.}~\bibnamefont
  {Fu}}\ and\ \bibinfo {author} {\bibfnamefont {E.}~\bibnamefont {Berg}},\
  }\href {\doibase 10.1103/PhysRevLett.105.097001} {\bibfield  {journal}
  {\bibinfo  {journal} {Phys. Rev. Lett.}\ }\textbf {\bibinfo {volume} {105}},\
  \bibinfo {pages} {097001} (\bibinfo {year} {2010})}\BibitemShut {NoStop}%
\bibitem [{\citenamefont {Sato}(2010)}]{Sato2010}%
  \BibitemOpen
  \bibfield  {author} {\bibinfo {author} {\bibfnamefont {M.}~\bibnamefont
  {Sato}},\ }\href {\doibase 10.1103/PhysRevB.81.220504} {\bibfield  {journal}
  {\bibinfo  {journal} {Phys. Rev. B}\ }\textbf {\bibinfo {volume} {81}},\
  \bibinfo {pages} {220504} (\bibinfo {year} {2010})}\BibitemShut {NoStop}%
\bibitem [{\citenamefont {Mukherjee}\ \emph {et~al.}(2017)\citenamefont
  {Mukherjee}, \citenamefont {Yan}, \citenamefont {Patel}, \citenamefont
  {Hadzibabic}, \citenamefont {Yefsah}, \citenamefont {Struck},\ and\
  \citenamefont {Zwierlein}}]{Mukherjee2016}%
  \BibitemOpen
  \bibfield  {author} {\bibinfo {author} {\bibfnamefont {B.}~\bibnamefont
  {Mukherjee}}, \bibinfo {author} {\bibfnamefont {Z.}~\bibnamefont {Yan}},
  \bibinfo {author} {\bibfnamefont {P.~B.}\ \bibnamefont {Patel}}, \bibinfo
  {author} {\bibfnamefont {Z.}~\bibnamefont {Hadzibabic}}, \bibinfo {author}
  {\bibfnamefont {T.}~\bibnamefont {Yefsah}}, \bibinfo {author} {\bibfnamefont
  {J.}~\bibnamefont {Struck}}, \ and\ \bibinfo {author} {\bibfnamefont {M.~W.}\
  \bibnamefont {Zwierlein}},\ }\href {\doibase 10.1103/PhysRevLett.118.123401}
  {\bibfield  {journal} {\bibinfo  {journal} {Phys. Rev. Lett.}\ }\textbf
  {\bibinfo {volume} {118}},\ \bibinfo {pages} {123401} (\bibinfo {year}
  {2017})}\BibitemShut {NoStop}%
\bibitem [{\citenamefont {Shin}\ \emph {et~al.}(2007)\citenamefont {Shin},
  \citenamefont {Schunck}, \citenamefont {Schirotzek},\ and\ \citenamefont
  {Ketterle}}]{Shin2007}%
  \BibitemOpen
  \bibfield  {author} {\bibinfo {author} {\bibfnamefont {Y.}~\bibnamefont
  {Shin}}, \bibinfo {author} {\bibfnamefont {C.~H.}\ \bibnamefont {Schunck}},
  \bibinfo {author} {\bibfnamefont {A.}~\bibnamefont {Schirotzek}}, \ and\
  \bibinfo {author} {\bibfnamefont {W.}~\bibnamefont {Ketterle}},\ }\href
  {\doibase 10.1103/PhysRevLett.99.090403} {\bibfield  {journal} {\bibinfo
  {journal} {Phys. Rev. Lett.}\ }\textbf {\bibinfo {volume} {99}},\ \bibinfo
  {pages} {090403} (\bibinfo {year} {2007})}\BibitemShut {NoStop}%
\bibitem [{\citenamefont {Kivelson}(1982)}]{Kivelson1982}%
  \BibitemOpen
  \bibfield  {author} {\bibinfo {author} {\bibfnamefont {S.}~\bibnamefont
  {Kivelson}},\ }\href {\doibase 10.1103/PhysRevB.26.4269} {\bibfield
  {journal} {\bibinfo  {journal} {Phys. Rev. B}\ }\textbf {\bibinfo {volume}
  {26}},\ \bibinfo {pages} {4269} (\bibinfo {year} {1982})}\BibitemShut
  {NoStop}%
\bibitem [{\citenamefont {Uehlinger}\ \emph {et~al.}(2013)\citenamefont
  {Uehlinger}, \citenamefont {Jotzu}, \citenamefont {Messer}, \citenamefont
  {Greif}, \citenamefont {Hofstetter}, \citenamefont {Bissbort},\ and\
  \citenamefont {Esslinger}}]{Uehlinger2013}%
  \BibitemOpen
  \bibfield  {author} {\bibinfo {author} {\bibfnamefont {T.}~\bibnamefont
  {Uehlinger}}, \bibinfo {author} {\bibfnamefont {G.}~\bibnamefont {Jotzu}},
  \bibinfo {author} {\bibfnamefont {M.}~\bibnamefont {Messer}}, \bibinfo
  {author} {\bibfnamefont {D.}~\bibnamefont {Greif}}, \bibinfo {author}
  {\bibfnamefont {W.}~\bibnamefont {Hofstetter}}, \bibinfo {author}
  {\bibfnamefont {U.}~\bibnamefont {Bissbort}}, \ and\ \bibinfo {author}
  {\bibfnamefont {T.}~\bibnamefont {Esslinger}},\ }\href {\doibase
  10.1103/PhysRevLett.111.185307} {\bibfield  {journal} {\bibinfo  {journal}
  {Phys. Rev. Lett.}\ }\textbf {\bibinfo {volume} {111}},\ \bibinfo {pages}
  {185307} (\bibinfo {year} {2013})}\BibitemShut {NoStop}%
\end{thebibliography}
\end{document}